\documentclass{aa}  
\usepackage{graphicx}
\usepackage{txfonts}
\usepackage{hyperref}
\usepackage{amsmath}
\usepackage[separate-uncertainty=true, multi-part-units=single]{siunitx}
\usepackage{comment}

\def\ltsima{$\; \buildrel < \over \sim \;$}
\def\simlt{\lower.5ex\hbox{\ltsima}}
\def\gtsima{$\; \buildrel > \over \sim \;$}
\def\simgt{\lower.5ex\hbox{\gtsima}}
\def\gsimeq
{\hbox{\raise0.5ex\hbox{$>\lower1.06ex\hbox{$\kern-1.07em{\sim}$}$}}}
\def\lsimeq
{\hbox{\raise0.5ex\hbox{$<\lower1.06ex\hbox{$\kern-1.07em{\sim}$}$}}}

\def\xmm{{\it XMM-Newton }}
\def\rosat{{\it ROSAT}}

\def\xmm{{\it XMM-Newton}}
\def\chandra{{\it Chandra}}
\def\suzaku{{\it Suzaku}}

\def\fermi{{\it Fermi}}

\def\erosita{{eROSITA}}

\def\srg{{\it SRG}}
\def\efeds{{eFEDS}}

\def\apj{ApJ}
\def\aj{AJ}
\def\mnras{MNRAS}
\def\aap{A\&A}
\def\aaps{A\&AS}
\def\apjl{ApJ}
\def\apjs{ApJS}
\def\araa{ARA\&A}
\def\pasj{PASJ}
\def\nat{Nature}

\def\procspie{Proc. SPIE}

\def\cs{C {\sc vi}}
\def\os{O {\sc vii}}
\def\oe{O {\sc viii}}

\begin{document}

   \title{Abundance and temperature of the outer hot circum-Galactic medium}
   \subtitle{The \srg/\erosita\ view of the soft X-ray background in the \efeds\ field}

   \author{G. Ponti\inst{1,2} 
   \and
   X. Zheng\inst{2}
   \and
   N. Locatelli\inst{2}
   \and
   S. Bianchi\inst{3}
   \and
   Y. Zhang\inst{2}
   \and
   K. Anastasopoulou\inst{1}
   \and
   J. Comparat\inst{2}
   \and
   K. Dennerl\inst{2}
   \and
   M. Freyberg\inst{2}
   \and
   F. Haberl\inst{2}
   \and
   A. Merloni\inst{2}
   \and
   T. H. Reiprich\inst{4}
   \and
   M. Salvato\inst{2}
   \and
   J. Sanders\inst{2}
   \and
   M. Sasaki\inst{5}
   \and
   A. Strong\inst{2}
   \and
   M. C. H. Yeung\inst{2}
   }
   \institute{INAF-Osservatorio Astronomico di Brera, Via E. Bianchi 46, I-23807 Merate (LC), Italy \\
        \email{gabriele.ponti@inaf.it}
    \and
    Max-Planck-Institut f{\"u}r extraterrestrische Physik, Giessenbachstrasse, D-85748, Garching, Germany 
    \and
    Dipartimento di Matematica e Fisica, Universit\`a degli Studi Roma Tre, via della Vasca Navale 84, I-00146 Roma Italy
    \and
    Argelander-Institut f\"ur Astronomie (AIfA), Universit\"at Bonn, Auf dem H\"ugel 71, 53121 Bonn, Germany
    \and
    Remeis Observatory and ECAP, Universit\"at Erlangen-N\"urnberg, Sternwartstrasse 7, D-96049 Bamberg, Germany
    }

   \authorrunning{G. Ponti et al.}

   \date{Received \today; accepted future}
 
  \abstract{Despite their vital importance to understand galaxy evolution and our own Galactic ecosystem, our knowledge of the physical properties of the hot, X-ray emitting, phase of the Milky Way is still inadequate.
  However, sensitive SRG/\erosita\ large area surveys are now providing us with the long sought-after data needed to mend this state of affairs. }
  {We aim to constrain the properties of the Milky Way hot halo emission towards intermediate Galactic latitudes close to the Galactic anti-center.}
  {We analyse the spectral properties of the integrated soft X-ray emission observed by \erosita\ in the relatively deep \efeds\ field.} 
  {We observe a flux of $12.6$ and $5.1\times10^{-12}$ erg cm$^{-2}$ s$^{-1}$ deg$^{-2}$ in the total (0.3--2) and soft (0.3--0.6~keV) band. 
  We measure the temperature and metal (Oxygen) abundance of the hot circum-Galactic medium (CGM) to be within $kT_{CGM}=0.153-0.178$~keV and $Z_{CGM}=0.052-0.072$~$Z_\odot$, depending on the contribution of solar wind charge exchange (SWCX). 
  Slightly larger CGM abundances $Z_{CGM}=0.05-0.10$~$Z_\odot$ are possible, considering the uncertain extrapolation of the extragalactic Cosmic X-ray background (CXB) emission below $\sim1$~keV. 
  To recover CGM abundances as large as $Z_{CGM}=0.3$~$Z_\odot$, it must be postulated the presence of an additional component, likely associated with the warm-hot intergalactic medium, providing $\sim15-20$~\% of the flux in the soft X-ray band.
  We observe line widths of the CGM plasma smaller than $\Delta v\leq500$ km s$^{-1}$.  \\
  The emission in the soft band is dominated ($\sim47$~\%) by the circum-Galactic medium (CGM), whose contribution reduces to $\sim30$~\% if Heliospheric SWCX contributes at the level of $\sim15$~\% also during solar minimum. 
  The remaining flux is provided by the CXB ($\sim33$~\%) and the local hot bubble ($\sim18$~\%).
  Moreover, the \erosita\ data require the presence of an additional component associated with the elusive Galactic corona plus a possible contribution from unresolved M dwarf stars. This component has a temperature of $kT\sim0.4-0.7$~keV, a considerable ($\sim$~kpc) scale-height and it might be out of thermal equilibrium. It contributes $\sim9$~\% to the total emission in the 0.6--2 keV band, therefore it is a likely candidate to produce part of the unresolved CXB flux observed in X-ray ultra-deep fields. 
  We also observe a significant contribution to the soft X-ray flux due to SWCX, during periods characterised by stronger solar wind activity, and causing the largest uncertainty on the determination of the CGM temperature. }
  {We constrain temperature, emission measure, abundances, thermal state, and spectral shape of the outer hot CGM of the Milky Way. }


   \maketitle
%

\section{Introduction} 

We are living in a golden age for Galactic astrophysics. 
On the one hand, the Gaia satellite, together with large spectroscopic surveys are allowing us to understand the dynamics and composition of the stars of the Milky Way to a degree never reached before (Gaia Collaboration 2016; 2021; Majewski et al. 2017). 
On the other hand, the standard cosmological model dictates that the formation and evolution of Milky Way-like galaxies is governed by the elusive dark matter halo (White \& Rees 1978; White \& Frenk 1991; Dodelson et al. 2004; Mo et al. 2010).
In particular, state of the art cosmological simulations suggest that the dominant component of galactic baryons in the present day Universe should reside in their halos, within the so-called circum-galactic medium (CGM; Crain et al. 2010; Tumlinson et al. 2017; Bogdan et al. 2015; Kelly et al. 2021; Oppenheimer et al. 2020; Truong et al 2020). 
Additionally, they predict that the growth and evolution of galaxies critically depends on the physics of the multi-phase inter-stellar medium (ISM) and CGM (Putman et al. 2012; Tumlinson et al. 2017; Naab et al. 2017). 
In particular, the latter is expected to be dominated by its hotter component, which is forming a rarefied plasma close to the virial temperature ($kT\sim0.15-0.2$~keV) and extending to the virial radius ($R\sim200$~kpc).  
Therefore, such plasma is expected to form a diffuse emission component, over the entire sky. 
Despite its vital importance, our knowledge of the hot Galactic plasma is still to be endeavoured. 

Since its discovery, the study of the Cosmic X-ray background (CXB) has been a major field of research (Giacconi et al. 1962). 
The CXB appears as a uniform X-ray glow over the entire sky, whose energy spectrum has been measured to be consistent with a power law with photon index of $\Gamma\sim1.45$ in the 2--10 keV band, which then breaks to a steeper slope, therefore creating a peak in the energy spectrum, around $\sim30$~keV and to roll over at higher energies (Marshall et al. 1980; Vecchi et al. 1999; Revnivtsev et al. 2003; 2005; De Luca \& Molendi 2004; Hickox \& Markevitch 2006; Kushino et al. 2002; Gilli et al. 2007). 
In particular, the advent of \xmm\ and \chandra\ has allowed us to make a giant leap forward in our understanding of the CXB, thanks to an array of extra-galactic surveys going from the ultra-deep ($\sim7$~Ms) pencil beam exposures (Luo et al. 2017) to much larger-area, but shallower, surveys (see Brandt et al. 2021 for a review). 

Such extra-galactic surveys revealed that the majority of the X-ray background above $\sim0.5$ keV is composed of a large number of faint distinct sources. 
Indeed, they allowed us to resolve more than $\sim80$~\% and $\sim92$~\% of the CXB flux into discrete sources (i.e., active galactic nuclei, clusters of galaxies, groups, normal galaxies, etc.) in the 0.5-2 and 2-7 keV band, respectively (Luo et al. 2017; Brandt et al. 2021). 
Instead, the X-ray background appears to be truly diffuse in the softest energies, below $\sim0.5$~keV. 

In the nineties, the all sky \rosat\ maps revolutionised our understanding of the X-ray background in the softer energy band (Snowden et al. 1991; 1994; 1995; 1997). In particular, the sensitive \rosat\ images have revealed that the soft X-ray background is highly in-homogeneous and filling the entire sky (Snowden et al. 1991; 1997).
The \rosat\ data allowed astronomers to disentangle the emission from the local hot bubble (LHB) from the Galactic-scale emission. The former component manifests itself as a hot ($kT\sim0.1$~keV) bubble surrounding the Sun and with a radius of $\sim200$~pc (Liu et al. 2017; Zucker et al. 2022), therefore dominating the Cosmic X-ray background (CXB) in the softest band ($E<0.2$~keV; Liu et al. 2017).
At energies between $\sim0.2-0.6$~keV, the Galactic-scale emission dominates over the LHB and the CXB\footnote{Hereinafter, to avoid confusion, we will refer with the term CXB to the extra-galactic component of the X-ray background, clearly separating it from the other constituents which become more relevant in the soft X-ray band. We will also consider models for the CXB which have a larger flux in the soft band (CXBs) and larger flux in the hard one (CXBh; see Sect. \ref{CXB} for more details). }. 
Such Galactic component was interpreted as either a Galactic corona, which would be produced by a thickened disc with a scale-height of few kpc, or as the emission from the hot halo, extending out to the virial radius ($r_v\sim200$~kpc). 
Unfortunately, the low energy resolution of the \rosat\ cameras did not allow astronomers to disentangle the emission lines from the thermal continuum. Therefore, accurate measurements of the temperature and abundances of such Galactic component was not feasible. 
Indeed, one of the major results of this work is the characterisation of the physical properties (i.e., temperature, emission measure, abundances, etc.) of the Galactic component. 
Another is to demonstrate that both the CGM\footnote{Within a galaxy the CGM would comprise both the coronal component, the halo as well as other possible constituents. Throughout this work, we will use the term "CGM" (and not "halo") to refer to all the contributions apart from the hotter Galactic corona. It is likely that the halo emission constitutes the dominant part of what we call CGM, however we prefer to avoid using the term halo because we still have to probe whether such component possesses a density distribution consistent with a halo component or whether this emission has a significant fraction produced by a warm corona. } and the Galactic corona components are required by the \erosita\ data. Finally, we reckon it is important that we verified that the a steep continuum (that we associate with hot baryons in the warm-hot intergalactic medium) can contribute with a flux of the order of less than $\sim10^{-12}$~erg~cm$^{-2}$~s$^{-1}$~deg$^{-2}$, in the 0.3-0.6~keV band, if the CGM abundances are high ($Z_{CGM}\gg0.1$~$Z_\odot$). 

Over the last two decades, by accumulating hundreds of \xmm\ and \suzaku\ observations, astronomers have tried to constrain the soft X-ray emission from the Milky Way (Henley et al. 2010; 2012; 2013; 2015; Miller \& Bregman 2013; 2015; 2016; Yoshino et al. 2009; Nakashima et al. 2018). Indeed, they have shown that the Galactic X-ray emission, outside of the \fermi\ bubbles, can be reproduced by either a beta model with $kT\sim0.2$~keV and an extension of several hundred kpc (with abundances assumed to be $Z_{CGM}=0.3$~$Z_\odot$; Miller \& Bregman 2016; Bregman et al. 2018) or by an exponential atmosphere with a scale-height of a few kiloparsecs (Yao et al. 2005; 2007; 2008; Wang et al. 2005; 2009; 2010). 
Additionally, studies of absorption lines imprinted on the spectrum of few tens of bright AGN and of the dispersion measure of fast radio bursts have provided further observational evidence for the presence of either a hot halo around the Milky Way or an exponential atmosphere (Fang et al 2015; Miller \& Bregman 2015; Prochaska et al. 2019). Some of the most recent results include both components, however they often find the halo component to be dominant (Bregman et al. 2018). 

Very instructive is the comparison of what is observed in nearby galaxies. Surprisingly, deep observations of single normal galaxies have often failed to detect an X-ray halo extending to the virial radius, such as supposed to be surrounding the Milky Way (Wang 2001; 2003; Anderson et al. 2011; 2016; Li \& Wang 2013a,b; Li et al. 2017). 
Indeed, the clearest detections of hot Galactic plasma were reported around edge-on massive spirals, where the hot plasma is observed to form a thick atmosphere, like a corona, extended several kiloparsecs above and below the disc (Anderson et al. 2011; 2016; Wang 2001; 2003). 
Only stacks of samples of galaxies allowed to reach the signal to noise required to detect the hot halo beyond several tens of kiloparsecs (Anderson et al. 2015; Li et al. 2018; Comparat et al. 2022).

In addition to the intrinsic challenges associated with the detection of such faint and extended galactic haloes, an additional complication is typically affecting the soft X-ray band. 
Indeed, it has been demonstrated that the process of charge exchange between the ionised particles of the Solar wind with neutrals within the Heliosphere can induce a time-variable component to the soft X-ray background, which can be significantly brighter than such galactic haloes (Snowden et al. 2004; Kuntz 2019). 

With more than 3 million photons in the soft X-ray band, the \erosita\ (Predehl et a. 2021) Final Equatorial Depth Survey (eFEDS) provides us with an unprecedented possibility to study the characteristics of the soft X-ray background (Snowden et al. 1997). The survey, defined during the \erosita\ performance verification phase, comprises about 142 square degrees, observed to a uniform depth of $\sim2.2$~ks in 2019 (Brunner et al. 2021), and re-observed as part of the on-going all-sky survey program in 2020 and 2021. 

\section{Dataset and data reduction} 
\label{data}

\begin{figure}[t]
\vspace{-4.0 cm}

\hspace{-1.5cm}
\includegraphics[width=0.7\textwidth]{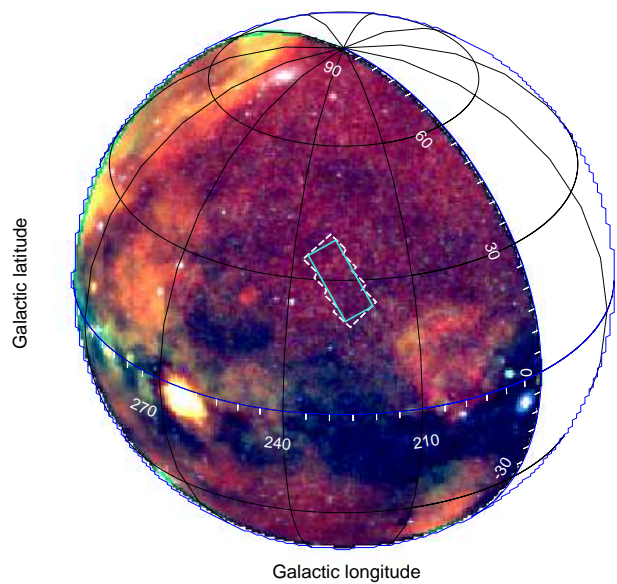}

\vspace{-4.0 cm}
\caption{Footprint of the \efeds\ field on the \erosita\ RGB sky map, as observed during e12. Red, green and blue colors show the emission in the 0.3-0.6, 0.6-1.0 and 1.0-2.3 keV energy bands. The radiation within the \efeds\ field is rather homogeneous and characteristic of the soft X-ray emission observed at latitudes away from the Galactic plane and away from the Galactic center. The dark stripe along the Galactic plane is primarily the byproduct of higher extinction there. 
}
\label{efedsF}
\end{figure}

We use both the public \erosita\ data from the \efeds\ field collected during the performance verification (sometimes abbreviated as PV) phase, which we refer to as e0, as well as the data from the same region accumulated during the first three passes of the all sky survey, which we will refer to as e1, e2, and e3, respectively, while with e12, we refer to the sum of the data from the first two passes\footnote{For extension e123 refers to the sum of the first three passes, and so on. } (see Fig. \ref{efedsF}; Brunner et al. 2021). 

The \efeds\ field covers $\sim142$ square degrees, extending from Galactic latitude $l$ from $\sim220^\circ$ to $\sim235^\circ$ and from Galactic longitude $b$ from $\sim20^\circ$ to $\sim40^\circ$ (see Fig. \ref{efedsF}). 
To avoid possible complications occurring at the edges of the \efeds\ region, we focus our analysis on the rectangular cyan region shown in Fig. \ref{efedsF}, which spans 107.5 square degrees on the sky. 
A similar surface brightness and color is observed both within the \efeds\ field and in regions away from the Galactic plane and away from the Galactic center (Fig. \ref{efedsF}). This suggests that the diffuse X-ray emission from the \efeds\ field is likely characteristic of the diffuse emission away from the Galactic plane and center.
Therefore, it represents an excellent field to study the emission from the soft X-ray background, which is not impacted by the Galactic outflow (Sofue 2000; Bland-Hawthorn et al. 2003; Su et al. 2010; Ponti et al. 2019; 2021; Predehl et al. 2020).
In this work, we will consider the total emission from such region including: point sources; extended sources; diffuse emission; and background. 

We investigated the temporal evolution of the particle background during the \erosita\ observations of the \efeds\ field and we found, in e3, an instance (possibly associated with a coronal mass ejection) when the particle background is enhanced by $\sim80$~\% and $\sim30$~\% in the $2.3-4.5$ and $0.3-1.4$ keV bands, respectively. No such events are observed during e0, e1 or e2. In particular, Figure \ref{FlareGTI} shows the comparison between the e12 spectrum applying the filtering for background flares (with the {\sc flareGTI} tool) and not. The consistency between these spectra corroborate the fact that important background flares do not affect the e12 spectrum, therefore they do not have an effect on the results obtained here. 

The \efeds\ field was observed with an exposure depth of approximately $\sim2.2$~ks during the PV phase (e0), while for about $\sim250$~s during each of the three all sky surveys. 
The \efeds\ field was scanned by \erosita\ during the periods: from the third to the ninth of November 2019 during e0; from the thirtieth of April to the fourteenth of May 2020 during e1; from the first to the fourteenth of November 2020 during e2; and from the third to the fifteenth of May 2021 during e3. 
We reduced the data with the eSASS Software version 947 (Brunner et al. 2021). 
In particular, we utilised the users release eSASSusers\_201009 from October 2020, which has significant improvements in the energy calibrations of each camera (Dennerl et al. 2020). 
We considered only single and double events. 

To avoid contamination from light leak (Predehl et al. 2021), which affects the cameras TM5 and TM7, we use only the "on-chip" filter cameras, which are: TM1, TM2, TM3, TM4 and TM6. 

We fit all spectra with the {\sc Xspec} software (Arnaud 1996) version 12.11.1. 
Uncertainties are reported at the 1-$\sigma$ confidence level for one interesting parameter, unless otherwise stated. We use the $\chi^2$ statistics to fit separately the different TM cameras aboard \erosita\ (although tying all parameters of the model reproducing the emission from the sky), however we show the combined spectra and residuals for display purposes only. We assume the Lodders (2003) abundances and Verner et al. (1996) cross sections, unless stated otherwise.

\section{Absorption}
\label{SectNH}

We estimated the column density of Galactic neutral material from the HI4PI data-cube\footnote{The HI4PI data traces only neutral hydrogen, therefore it should be considered as a lower limit to the real column density of absorbing material, being the ionized and molecular hydrogen unaccounted for. However, for column densities as low as the ones towards the \efeds\ field, the contribution from these components is typically observed to be small (see Fig. 13 of Schellenberger et al. 2015).} (HI4PI Collaboration et al. 2016). We divided the $\sim107.5$ square degrees of the \efeds\ field that we analysed into pixels of 11.7 square arcminutes (0.00325 square degrees) and recorded the average column density in the HI4PI map within such pixels. 
By approximating the observed distribution with a log-normal, we measured a mean column density of neutral Hydrogen of log$(N_H/$cm$^{-2})=20.51$ and standard deviation $\sigma_{\log(N_{\rm H}/{\rm cm^{-2}})}=0.117$.  
Hereinafter, following Locatelli et al. (2022), we reproduced the effects of the observed distribution of the neutral absorption column densities by employing the {\it disnht} model.

\section{Impact of the instrumental background}
\label{SectBack}

\begin{figure}[t]
\hspace{-0.9cm}
\vspace{-0.4cm}
\includegraphics[width=0.58\textwidth]{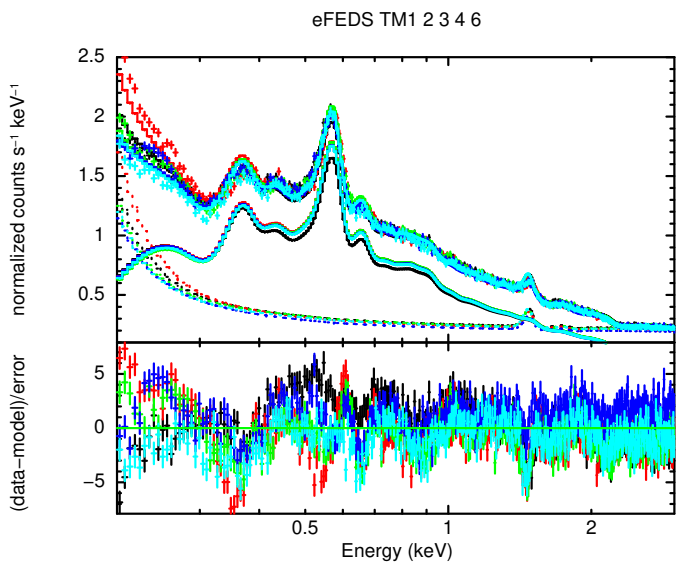}
\caption{Total X-ray emission from the eFEDS field as observed by \erosita. The spectra contain emission from all point-like and extended sources, diffuse emission from the sky as well as instrumental background. Black, red, green, blue and cyan data display the spectra observed with the TM1, TM2, TM3, TM4 and TM6 cameras aboard \erosita, respectively. The solid lines show a model aiming at reproducing the integrated emission from the sky (point-like and extended sources plus diffuse emission), while the dotted lines show the model for the instrumental background, derived from the filter wheel closed data (Freyberg et al. 2020). }
\label{spec}
\end{figure}

Figure \ref{spec} shows the total emission from the \efeds\ field, including all point and extended sources, diffuse emission from the sky as well as instrumental background. 
Thanks to the analysis of the filter wheel closed data, the \erosita\ team has developed a model of the instrumental background for each camera aboard \erosita. Such models are shown by the dotted lines in Fig. \ref{spec} (see also Freyberg et al. 2020). 
At energies above $\sim0.5-1$~keV, the instrumental background is dominated by a flat power law with a photon index close to $\Gamma=0$, plus a series of emission lines, which are induced by interactions of particles with the detectors and other components of the satellite. 
At energies below $\sim0.5$~keV, an increase in the background level is observed, as a consequence of electronic noise (see the steep power law shape dominating at low energy). 

We observe that the instrumental background dominates over the sky emission at energies below $E~\lsimeq~0.25$~keV and above $E~\gsimeq~1.5$~keV (Fig. \ref{spec}). 
We first perform a fit of the spectrum of each camera over the entire energy range from 0.2 to 3 keV.
We use a different instrumental background model for each camera, as specified by the \erosita\ team (Freyberg et al. 2020). 
All parameters of the instrumental background model are fixed, apart from a normalisation factor which is adjusted, for each camera, to the value necessary to fit the data above 2 keV, where the background is almost a factor of $\sim5-10$ stronger than the emission from the sky\footnote{This is a consequence of the drop in effective area above $\sim2$~keV.}. 
Then, for each camera, we fix the normalisation of the internal background model to the observed best fit parameter and we subsequently leave it fixed to such value. 
We point out that such technique is able to adjust for the variations in the rate of particles inducing hard X-ray emission, however it is not suited for determining the noise component below ~0.25 keV, which is caused by different effects and thus not strictly correlated with the high energy background. 
Indeed, we observe large residuals below $\sim0.3$~keV (Fig. \ref{spec}). 
Therefore, considering the still limited knowledge of the instrumental background and its evolution with time, we decided to fit the spectrum only within 0.3 and 1.4~keV, to reduce the impact on our best fit models of possible variations of the instrumental background. 

\section{Solar Wind Charge Exchange (SWCX)}
\label{SectSWCX}

The interaction of the ionised particles of the Solar wind with the flow of neutral ISM which constantly passes through the Heliosphere produces diffuse soft X-ray emission by charge exchange (Snowden et al. 2004; Kuntz 2019).
The brightness of such component is expected and observed to be modulated by the properties of the Solar wind, therefore to be variable over time. 
Thanks to the scanning strategy of \erosita, we can probe all these time-scales, therefore verifying the impact of any variable components on the observed emission. 

\subsection{Variations induced by SWCX}

\begin{figure*}[th]
\centering
\includegraphics[width=0.48\textwidth]{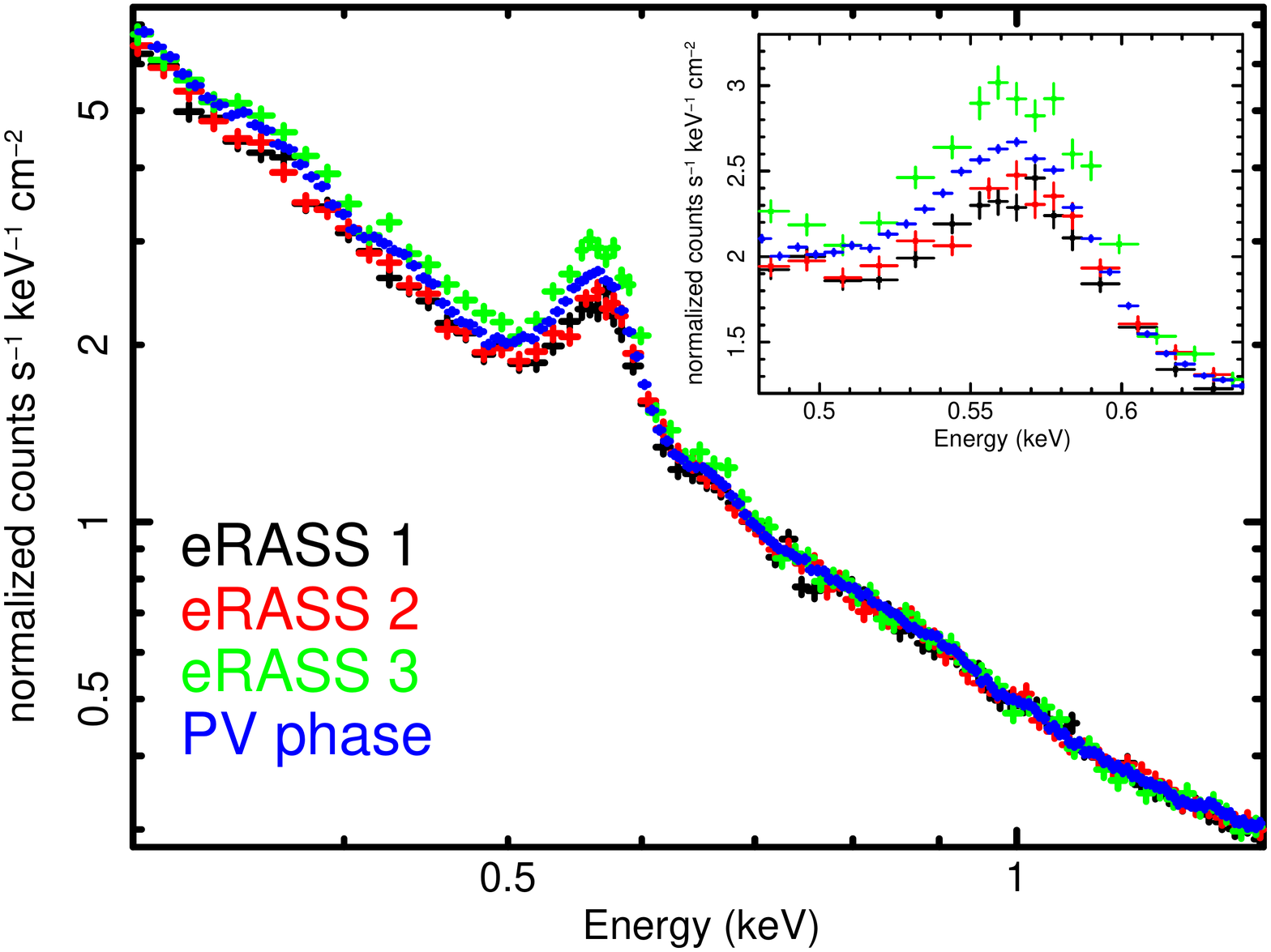}
\includegraphics[width=0.51\textwidth]{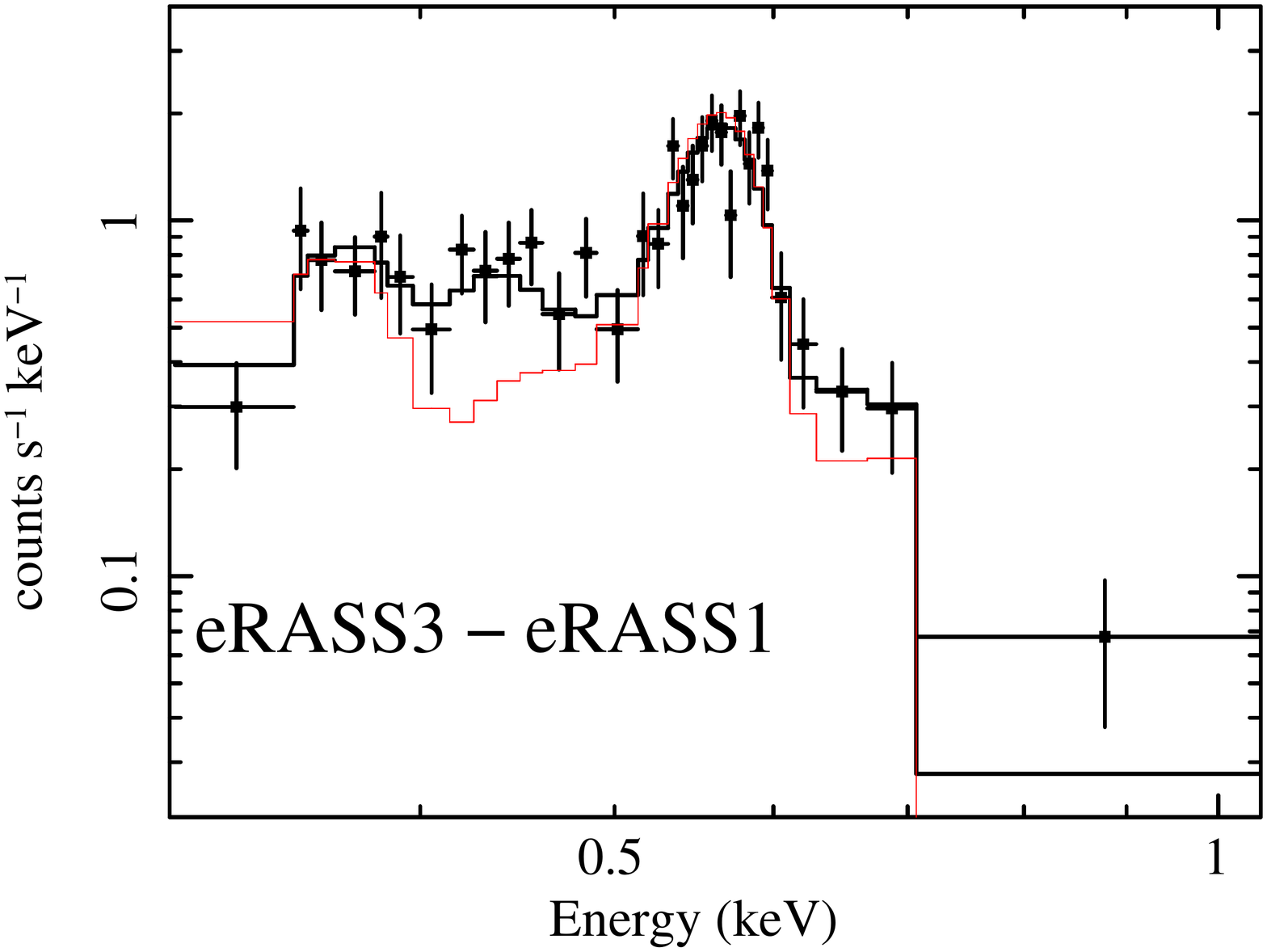}
\caption{{\it (Left panel)} Diffuse X-ray emission as observed by \erosita\ in the \efeds\ field during e1, e2, e3 and e0 (eRASS1, eRASS2, eRASS3 and the PV phase observations) in black, red, green and blue, respectively (the emission from different TMs is combined for display purposes). 
An enhancement at the energies of the soft X-ray emission lines is observed during e3, compared with e1 and e2. 
Such enhancement is characterised by a spectral shape which is characteristic of the emission induced by SWCX. 
A similar enhancement is observed during e0. The inset shows an enlargement of the spectra around the \os\ line, where the enhancement is most evident. 
{\it (Right panel)} Spectrum of the variable diffuse emission component fitted with a model for the SWCX. The black line shows the best fit model, while the red line shows a different implementation of the same model. }
\label{FSWCX}
\end{figure*}

The black, red, green and blue data in the left panel of Fig. \ref{FSWCX} show the total X-ray emission (including: point sources; extended sources; diffuse emission; and background) as observed by \erosita\ in the \efeds\ field during e1, e2, e3 and e0 (the PV phase observations), respectively\footnote{Hereinafter, we will refer to the total emission as "diffuse". This is justified by the fact that the diffuse emission dominates over all other contributions on the large scales considered here. Indeed, we observe that, within the $\sim0.3-2$~keV band, on scales of a fraction of a square degree or larger, the total emission is dominated by the diffuse component (including truly diffuse hot plasma components as well as the integrated emission from point sources, such as the CXB and the Galactic ridge emission). }. The diffuse emission has a low value and it is constant during e1 and e2. 
Indeed, the data points of the e1 and e2 spectra are consistent with each other over the $\sim0.3-1.4$~keV energy band (Fig. \ref{FSWCX}). 
On the other hand, enhanced emission is observed between $\sim0.3-0.7$~keV during e3 and e0.  
In particular, the peak flux of the \os\ line increases from $\sim2.6$ to $\sim3.3$~ph s$^{-1}$ keV$^{-1}$ cm$^{-2}$, corresponding to an increase of the order of $\sim25$~\% during e3 (see inset of Fig. \ref{FSWCX}). 
For this reason, hereinafter, we will consider primarily the data taken during e1 and e2. 

To determine the total \os\ and \oe\ line intensities observed during e12, we fitted the e12 spectrum, within a narrow energy band (0.45--0.75~keV), with an absorbed power law and two narrow Gaussian lines, plus the instrumental background\footnote{The absorption is fixed to the values discussed in Sect. \ref{SectNH}, while the background to the values discussed in Sect. \ref{SectBack}.}. 
We obtain best fit intensities for the \os\ and \oe\ of $I_{\os}=3.1\pm0.1$ ph s$^{-1}$ cm$^{-2}$ sr$^{-1}$ and $I_{\oe}=0.28\pm0.03$ ph s$^{-1}$ cm$^{-2}$ sr$^{-1}$, respectively. 

\subsection{Constraining SWCX from the difference spectrum}
\label{SectDiffSWCX}

The right panel of Fig. \ref{FSWCX} shows the spectrum of the variable component, obtained from the difference of the spectrum observed during e3 minus the one observed during e1. In this way, all constant components are subtracted, leaving only the variable emission. 

We fit the difference spectrum with a solar wind charge exchange model ({\sc acx2} model in {\sc xspec}, which is part of the {\sc atomdb} package; Smith et al. 2012; Foster et al. 2020). 
We assume solar abundances\footnote{The {\sc acx2} model assumes the Anders \& Grevesse (1989) abundances.} and single recombination\footnote{We also assume the acx model number four, for more details on this assumption please see the model documentation at: https://acx2.readthedocs.io/en/latest/.}. Additionally, we assume a collision speed velocity of $450$~km~s$^{-1}$, in order to reflect the solar wind speed, however we try two different implementations of this velocity. 
The first attempt assumes that $450$~km~s$^{-1}$ corresponds to the center of mass velocity, while the second trial assumes that it corresponds to the donor ion velocity. 
The right panel of Fig. \ref{FSWCX} shows that an acceptable fit can be obtained with the first attempt (black line), while the second trial leaves unacceptable residuals. 
Therefore, hereinafter we assume a collision speed velocity of $450$~km~s$^{-1}$, corresponding to the center of mass velocity. 

The best fit plasma temperature of the ionised component and the fraction of neutral Helium result to be $kT=0.136\pm0.007$ keV and $F_{He0}>0.2$, with a normalisation of $0.25\pm0.09$, corresponding to an average flux of $7.4\times10^{-11}$~erg~cm$^{-2}$~s$^{-1}$ over the \efeds\ area in the 0.4-0.6~keV band, which correspond to $6.9\times10^{-13}$~erg~cm$^{-2}$~s$^{-1}$~deg$^{-2}$ ($\chi^2=168.4$ for 161 dof; see Tab. \ref{TSWCX}).
On the one hand, we note that such values are consistent with this emission being produced by solar wind charge exchange. 
On the other hand, we note that the spectrum is composed of the variable component only (missing the emission from the constant one), therefore it is most likely that the best fit values are biased, hampering us from going into a deeper investigation of this effect in this work. 
\begin{table}
\centering
    \caption{Best fit parameters of the variable component of the emission observed in the \efeds\ field, once the difference spectrum (e3 minus e1) is fitted with the {\rm acx2} model. 
    The first column ($kT_{SWCX}$) reports the plasma temperature, in keV. 
    The second column ($F_{He0}$) reports the fraction of neutral Helium (relative to the total neutral population, assumed to be H and He) in the plasma. The third column reports the flux of the acx2 component in the 0.4-0.6~keV band, over the \efeds\ region, in units of $10^{-13}$ erg cm$^{-2}$ s$^{-1}$ deg$^{-2}$. 
    Then the $\chi^2$ and the degrees of freedom are reported. }
    \label{TSWCX}
    \begin{tabular}{c c c c c c c c c c c c }
    \hline \hline
\multicolumn{5}{c}{SWCX}         \\
    \hline \hline
$kT_{SWCX}$     & $F_{He0}$          & Flux & $\chi^2$     & $dof$        \\ 
$0.137\pm0.004$ & $>0.2$             & $6.9\pm1.5$ & 168.4        & 161              \\
\hline \hline
    \end{tabular}
\end{table}

\subsection{Spatial distribution of the excess emission and association to Heliospheric SWCX}

We investigated the map, in the 0.5-0.6 keV band, of the emission observed during e3, which is characterised by a more intense solar wind. 
From the map accumulated during e3, we have subtracted the emission in the same band accumulated during e12. 
Such difference map is consistent with a constant excess over the entire region. 
This confirms that the SWCX emission is associated with an increased glow over the entire \efeds\ region (see Ponti et al. subm.), indicating that the excess of SWCX emission occurs on time-scales longer than $\sim6$ days, which is the time it took to scan the \efeds\ region. 

Such behaviour appears to be remarkably different from the variability pattern observed in \xmm\ and \chandra\ data, where the variations induced by SWCX occur on hours-days time-scales. 
It is likely that the high Earth orbit of \xmm\ and \chandra\ make them more sensitive to the rapidly variable SWCX emission occurring at the edge of the Earth magnetosphere (Snowden et al. 2004; Kuntz 2019), while for the orbit around L2 of \erosita\ this component is missing, so that \erosita\ is sensitive to the more-slowly varying Heliospheric component of the SWCX emission (Kuntz 2019; Dennerl et al. in prep.).

\subsection{SWCX emission during e12}

Our analysis demonstrates that SWCX is present during e0 and e3, while it still remains to be demonstrated whether SWCX is present also during e1 and e2. Theoretical arguments suggest that such component must be present also at solar minimum, although at a lower level. 

Currently, various attempts are in progress to establish what is the contribution to the total emission due to SWCX during e1 and e2. 
\begin{itemize}
\item{} Through a preliminary investigation of the evolution in time of the SWCX component, Dennerl et al. (in prep.) are estimating a count rate in the 0.4-0.6 keV band of $\sim0.068$, $\sim0.074$, and $\sim0.16$~ph~s$^{-1}$~cm$^{-2}$ within the \efeds\ field during e1, e2, and e3, respectively. These estimates are consistent with the flux of the SWCX component observed by analysing the e3 minus e1 spectrum (with a count rate of $0.09$~ph~s$^{-1}$~cm$^{-2}$ and flux $(6.9\pm1.5)\times10^{-13}$~erg~cm$^{-2}$~s$^{-1}$ deg$^{-2}$ in the 0.4-0.6 keV band; see Tab. \ref{TSWCX}). Assuming the SWCX spectrum observed in Sect. \ref{SectDiffSWCX}, these values correspond to a flux of $\sim1.2$ ph s$^{-1}$ cm$^{-2}$ sr$^{-1}$ in the \os\ line during e12.

\item{} Through the study of nearby Ophiuchus dark cloud, Yeung et al. (in prep.) are preliminary estimating the flux of the SWCX component observed by \erosita. 
Yeung et al. (in prep.) observe a flux of F$_{SWCX}=(2.1\pm0.6)\times10^{-13}$ and F$_{SWCX}=(6.1\pm0.7)\times10^{-13}$ erg cm$^{-2}$ s$^{-1}$ deg$^{-2}$ in the 0.4-0.6~keV band towards the Ophiuchus cloud during e1 and e2, respectively. 
These estimates are $\sim1.7$ times smaller than the previous ones and they might be the consequence of the different lines of sight, of the different times of observation\footnote{\erosita scans through the Ophiuchus cloud region about one month before the \efeds\ field.}, etc. Alternatively, they might reflect the intrinsic uncertainty on determining the intensity of the SWCX component. 
Considered the different SWCX model assumed by Yeung et al. (in prep.), the above 0.4-0.6~keV surface brightness translates into a flux of $\leq1.0$ ph s$^{-1}$ cm$^{-2}$ sr$^{-1}$ in the \os\ line during e12. 

\item{} Qu et al. (2022) have studied the temporal variation of the \os\ and \oe\ flux, as observed by \xmm, over a solar cycle. They found a significant variation induced by the Heliospheric SWCX component, which shows a minimum close to the solar minimum. 
In particular, they estimated the true Galactic \os\ and \oe\ emission lines to have mean values of the order of $\sim5.4$ and $1.7$ ph s$^{-1}$ cm$^{-2}$ sr$^{-1}$, respectively. 
We note that such values are about a factor of $\sim1.7$ and $\sim6$ times larger than the line intensities towards the \efeds\ field. 
This is in line with the fact that we are looking towards a line of sight away from the Galactic outflow and corroborates the fact that in the e12 spectrum the effects of SWCX are minimal. 
Additionally, we note that the \os\ flux measured by Qu et al. (2022) close to solar minimum are consistent with being entirely due to the Galactic \os\ emission. 
This, therefore, suggests a negligible contribution due to Heliospheric SWCX during solar minimum, such as during e12. 
\end{itemize}

We take the differences between these preliminary estimates of the normalisation of the SWCX component during e12 as a measure of the current uncertainty on its contribution, which goes from a negligible fraction to a flux of $\sim1.2$ ph s$^{-1}$ cm$^{-2}$ sr$^{-1}$ in the \os\ line.

\section{Definition of the initial model}

In this section we spell out all ingredients which compose the initial model of the spectrum and will be considered in all following fits. 

To minimise the contribution of the emission from SWCX, we fit the e12 spectra only. 
Indeed, despite the e0 spectrum has better signal to noise, the large and not-well-tracked fluctuations of the flux of the various ions composing the solar wind would induce significant systematic uncertainties to the best fit of e0 and e3. 

\subsection{SWCX}

We assume that the emission from SWCX is subdominant during e12. 
This is corroborated by the observation that: i) the spectra observed during e1 and e2 are consistent with each other; ii) the energy of the \os\ triplet is shifted towards the resonant line (Sect. 8.1); iii) the lines and continuum are consistent with being produced by an optically thin thermal plasma. 
Indeed, SWCX is expected to be: i) intrinsically time-variable (as a consequence of the variability of the solar wind, among other effects; Dennerl et al. in prep.); ii) is characterised by \os\ triplets dominated by the forbidden line; iii) has a continuum different from a bremsstrahlung; therefore, it is unlikely to provide a dominant contribution during e1 and e2. 

On the other hand, to quantify how our ignorance on the flux of the SWCX component during e12 propagates into our best fit results, we perform two sets of fits. One where we assume that the SWCX emission can be completely neglected. The second one assumes a SWCX component with an intensity as large as estimated by Dennerl et al. (in prep.), therefore corresponding to a flux of  $(6.9\pm1.5)\times10^{-13}$~erg~cm$^{-2}$~s$^{-1}$ deg$^{-2}$ in the 0.4-0.6~keV band and a spectral shape as constrained by the e3-e1 difference spectrum (Tab. \ref{TSWCX}; Sect. \ref{SectDiffSWCX}). 

\subsection{Cosmic X-ray Background (CXB)}
\label{CXB} 

\begin{figure}[t]
\centering
\includegraphics[width=0.5\textwidth]{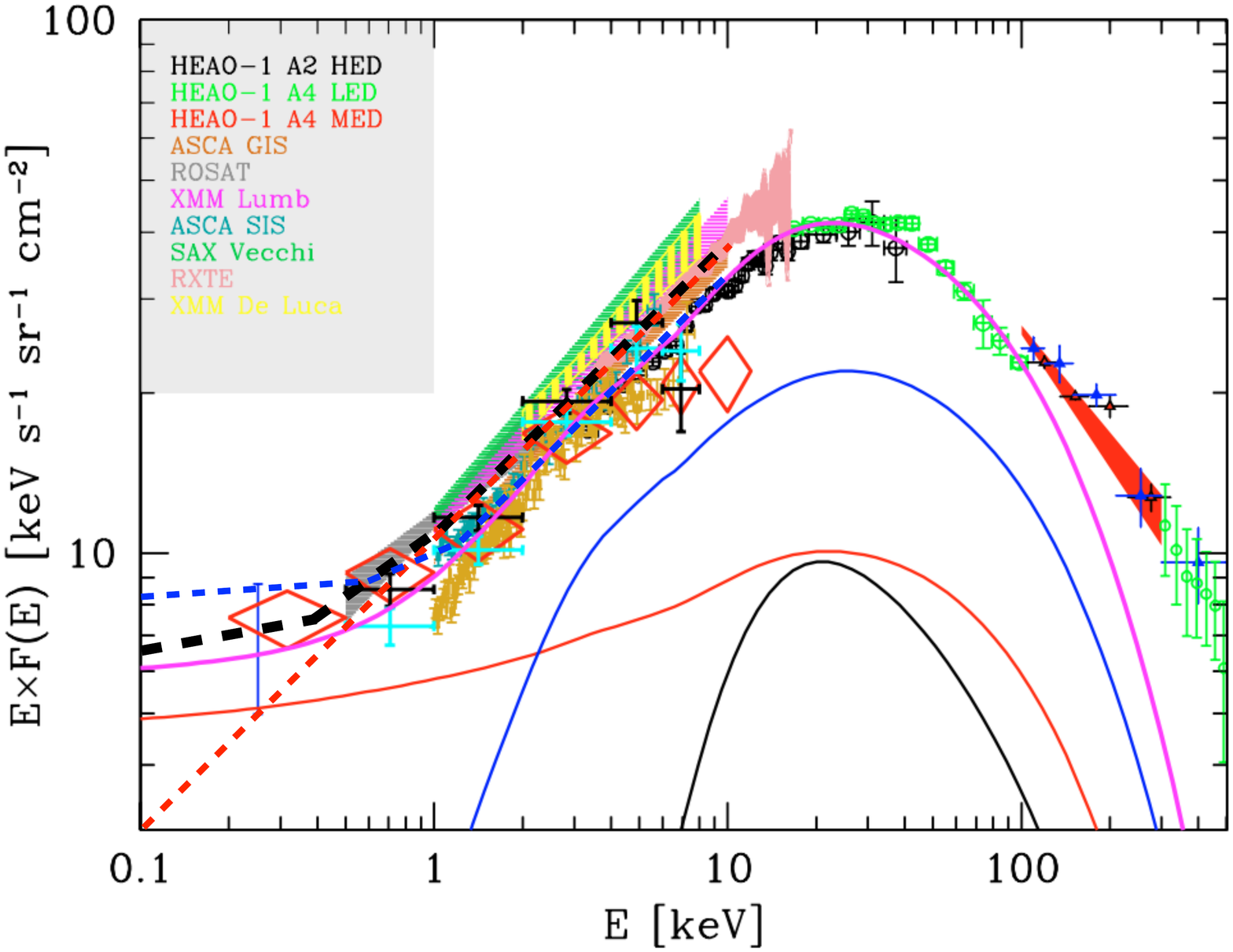}
\caption{The figure is taken from Gilli et al. (2007) which shows the CXB spectrum as observed by different instruments (as explained in the top left). The solid lines show the predicted contribution from the different components: the black, blue, red and magenta lines show the emission from Compton-thick, obscured Compton-thin, unobscured AGN, and total AGN plus galaxy cluster emission, respectively. 
The red dashed line shows the simplified CXB model often assumed in literature, composed of a power law shape with photon index $\Gamma=1.45$ and normalisation of 10.5 photons s$^{-1}$ cm$^{-2}$ sr$^{-1}$ at 1 keV (referred to CXBh in this work). In particular, such model fails to properly reproduce the data below $\sim1$~keV. 
The black dashed line shows the double broken power law model which has been defined in order to better reproduce the constraints on the cosmic X-ray background emission below $\sim1$~keV (referred to CXB in this work). 
It is composed of a power law with photon index of $\Gamma_1=1.9$ below 0.4 keV, then $\Gamma_2=1.6$ below 1.2 keV and then $\Gamma_3=1.45$ above 1.2 keV, with a normalisation of 8.2 photons s$^{-1}$ cm$^{-2}$ sr$^{-1}$ at 1 keV.
The blue dashed line shows a double broken power law which has been chosen to maximise the Cosmic X-ray emission from extragalactic sources, which is composed of a power law with photon index of $\Gamma_1=1.96$ below 0.6 keV, then $\Gamma_2=1.75$ below 1.2 keV and then $\Gamma_3=1.45$ above 1.2 keV, with a normalisation of 8.5 photons s$^{-1}$ cm$^{-2}$ sr$^{-1}$ at 1 keV (referred to CXBs in this work).}
\label{FCXB}
\end{figure}

The integrated emission from the Cosmic X-ray background, as measured by different X-ray instruments is shown in Fig. \ref{FCXB} (figure taken from Gilli et al. 2007)\footnote{The y-axis in Fig. \ref{FCXB} reports the values $E\times F(E)$, where $F(E)$ shows the flux as a function of energy: $F(E)=E \times N_{ph}(E)$, where $N_{ph}(E)$ is the number of photons as a function of energy in units of s$^{-1}$ cm$^{-2}$ sr$^{-1}$ keV$^{-1}$.}. 
Over the $\sim1$ to $\sim10$~keV energy range, the CXB can be described with a simple absorbed power law model ({\sc powerlaw} model in {\sc Xspec}) with photon index fixed to $\Gamma = 1.45$ and a normalisation of $\sim10.5$ ph cm$^{-2}$ s$^{-1}$ sr$^{-1}$ at 1 keV (see red dotted line in Fig. \ref{FCXB})\footnote{Fluctuations around the average normalisation of the CXB are expected to be observed as a consequence of the Cosmic variance.}. 
We will refer to this model component as CXBh. 
Additionally, considered its extra-Galactic nature, we assumed that the CXB is absorbed by the full column density of Galactic absorption. 
Detailed studies with \chandra, \xmm, \rosat\ and other X-ray instruments have resolved more than $\sim95$ \% of this component into point sources (AGN) and galaxy clusters, in the $2-8$~keV energy range (Hasinger et al. 1993; Hickox \& Markevitch 2006; Revnivtsev et al. 2003; Liu et al 2017). 
We note that the CXB normalization of $\sim10.5$ ph cm$^{-2}$ s$^{-1}$ sr$^{-1}$ at 1 keV for a photon index of $\Gamma=1.45$ corresponds to a normalisation of $\sim0.34$ ph cm$^{-2}$ s$^{-1}$ at 1~keV within the \efeds\ field of $\sim107.5$ square degrees. 
Unless otherwise stated, we leave the normalization of the CXB component to be free in the fits and we will verify a posteriori whether the normalisation of the CXB is within the range allowed by the Cosmic variance. 

We note that the CXB synthesis models containing also the contribution from groups and clusters, suggest that there might be a steepening of the CXB slope at energies below $\sim1$~keV (Gilli et al. 2007). 
To reproduce such steepening we considered a double broken power law, which is assumed to be identical to the simple power law model above 1.2 keV, but producing a higher flux at lower energies (black dashed line in Fig. \ref{FCXB}). 
The double broken power law has a photon index of $\Gamma_1=1.9$ below 0.4 keV, then $\Gamma_2=1.6$ between 0.4 and 1.2 keV and then $\Gamma_3=1.45$ above 1.2 keV, with a normalisation of 8.2 photons s$^{-1}$ cm$^{-2}$ sr$^{-1}$ at 1 keV (corresponding to 0.269 photons s$^{-1}$ cm$^{-2}$ at 1 keV over the \efeds\ area).
This appears as the most realistic representation of the constraints on the CXB accumulated so far (Gilli et al. 2007), therefore we will use such component (which we will refer to as CXB) in all our fits, unless stated otherwise. 

We also note that the observed data show a large scatter below $\sim1$~keV (Fig. \ref{FCXB}). 
Therefore, we define a third model which is still in rough agreement with the observational data and it maximises the emission in the soft X-ray band (see blue dashed line in Fig. \ref{FCXB}). 
The model has a double broken power law shape with a photon index of $\Gamma_1=1.96$ below 0.6 keV, then $\Gamma_2=1.75$ between 0.6 and 1.2 keV and then $\Gamma_3=1.45$ above 1.2 keV, with a normalisation of 8.5 photons s$^{-1}$ cm$^{-2}$ sr$^{-1}$ at 1 keV (black dashed line in Fig. \ref{FCXB}).
We will refer to this model as CXBs and we will employ it (alongside CXBh) in Sect. \ref{SectCGM} in order to understand how our assumptions on the CXB might systematically impact our results. 

\subsection{Local Hot Bubble (LHB)}

The Sun is located within a bubble of hot plasma possessing a temperature of $kT\sim0.1$ keV and an extension of $\sim10^2$ parsecs, which fills the local cavity (therefore it is un-absorbed in the X-ray band). Such bubble is typically called "local hot bubble" (Cox \& Snowden 1986; Snowden et al. 1990; Galeazzi et al. 2014; Liu et al. 2017).
Following the work of Liu et al. (2017), we assume that the emission from the LHB is well reproduced by a hot plasma component in thermal equilibrium ({\sc apec} model in {\sc Xspec}) with a temperature of $kT=0.097$~keV. Additionally, we assume that such component is un-absorbed, being located within only few $10^2$ parsecs from the Sun. Finally, we assume that it posses solar abundances. 
From Fig. 6 of Liu et al. (2017), we estimate the average emission measure associated with the LHB in the \efeds\ field, which results to be 0.00266 cm$^{-6}$ pc. 
The {\sc apec} normalisation $N_{apec}$ is defined as $N_{apec}=10^{-14}/(4\pi D_A^2) \times \int n_e n_H dV$, where $n_e$ and $n_H$ are the electron and Hydrogen densities (in cm$^{-3}$), respectively, $V$ is the volume (cm$^{-3}$) and $D_A$ is the angular diameter distance (in cm; see Xspec User Manual\footnote{https://heasarc.gsfc.nasa.gov/xanadu/xspec/manual/XspecManual.html}). This can be written as $N_{apec}=10^{-14} \frac{\int n_e n_H \times A \times dl}{4\pi D_A^2} = 10^{-14} \frac{\theta}{4 \pi} \int n_e n_H \times dl$, where $A$ is the section of the volume $V$ perpendicular to the line of sight and whose extension along the line of sight is $l$, and $\theta$ is the subtended solid angle (in steradians). 
Considering the projected area in the sky of the \efeds\ field, which is of $\sim107.5$ square degrees, corresponding to 0.0327 sr, this becomes: $N_{apec}= 10^{-14} \frac{0.0327}{4 \pi} 3.09\times10^{18} \times EM$[pc cm$^{-6}$]~$ \sim80\times EM$[pc cm$^{-6}$]. 
Hereinafter, we will convert all best fit {\sc apec} normalisations from Xspec into N$_{apec}$[pc~cm$^{-6}$]. 
As for the CXB, we leave the normalisation of the LHB free to vary and we check a posteriori whether its value agrees with the one observed by \rosat\ (Liu et al. 2017). 

\subsection{Circum Galactic medium (CGM)}

Finally, we assume that the emission from the circum Galactic medium is composed by hot plasma in thermal equilibrium ({\sc apec} model in {\sc Xspec}). 

\section{Detection of the elusive Galactic corona}
\label{SectCoro}

We start our investigation by fitting the \efeds\ spectrum with three components: two thermal components ({\sc apec} models in {\sc Xspec}) of which one for the LHB (red line in Fig. \ref{FCoro}) and one for the CGM (blue line in Fig. \ref{FCoro}) and a doubly broken power law for the CXB (magenta line in Fig. \ref{FCoro}), in addition to the instrumental background (black line in Fig. \ref{FCoro} and Tab. \ref{TAbu}) and, when specified, the SWCX emission (cyan line in the right panels of Fig. \ref{FCoro}). 
We note that, as a consequence of our assumptions, the spectral shape of the LHB and the CXB are fixed, therefore only their normalisations are allowed to vary in the fit. 

We first investigate the effects of assuming different sets of abundances (see Sect. \ref{SecAbu}) and decide to assume the solar abundances measured by Lodders et al. (2003).

\begin{table*}
\small
\centering
    \caption{Best fit parameters obtained by fitting the \efeds\ e12 spectrum with different models.
    LHB, CGM, CXB, Coro, Coro2, SWCX and shift stand for the local hot bubble, circum-Galactic medium, coronal emission in thermal equilibrium and out of thermal equilibrium, solar wind charge exchange and shift of the energy scales, respectively. 
    Each column shows the best fit parameters obtained for each model. Each adjacent pair of columns show the best fit results with the same model, under the assumption of either negligible or high SWCX contribution, respectively. 
    $N_{LHB}$, $N_{CGM}$, $N_{Coro}$ show the normalisations of the local hot bubble, circum-Galactic medium and coronal components, respectively, in units of $10^{-3}$ pc cm$^{-6}$. 
    $N_{CXB}$ shows the normalisation of the CXB component in units of photons keV$^{-1}$ cm$^{-2}$ s$^{-1}$ at 1 keV. 
    $kT_{CGM}$ and $kT_{Coro}$ show the temperatures of the circum-Galactic medium and coronal components in keV.
    $\tau$ shows the ionisation time-scales of the coronal plasma in units of $10^{10}$ s cm$^{-3}$. 
    }
    \label{TCoro}
    \begin{tabular}{c | c c | c c | c c | c c }
    \hline \hline
    \multicolumn{9}{c}{\bf SPECTRUM e12} \\
    \hline \hline
             &                   &                   &                   &                   & \multicolumn{2}{c}{shift}             & \multicolumn{2}{c}{shift}             \\
             & \multicolumn{2}{c}{LHB-CGM-CXB}       & \multicolumn{2}{c}{LHB-CGM-Coro-CXB}  & \multicolumn{2}{c}{LHB-CGM-Coro-CXB}  & \multicolumn{2}{c}{LHB-CGM-Coro2-CXB}   \\
             &                   &  SWCX             &                   & SWCX              &                   & SWCX              &                     & SWCX            \\
$N_{LHB}$    & $5.4\pm0.5$       & $5.7\pm0.4$       & $3.7\pm0.4$       & $4.0\pm0.4$       & $3.4\pm0.5$       & $3.6\pm0.5$       & $3.2\pm0.5$         & $3.4\pm0.6$        \\
$N_{CXB}$    & $0.300\pm0.002$   & $0.292\pm0.003$   & $0.276\pm0.003$   & $0.274\pm0.003$   & $0.275\pm0.003$   & $0.273\pm0.003$   & $0.270\pm0.003$     & $0.268\pm0.003$     \\
$kT_{CGM}$   & $0.190\pm0.005$   & $0.240\pm0.006$   & $0.166\pm0.003$   & $0.191\pm0.006$   & $0.162\pm0.003$   & $0.181\pm0.005$   & $0.157\pm0.004$     & $0.173\pm0.005$     \\
$Z_{CGM}$    & $0.085\pm0.007$   & $0.12\pm0.02$     & $0.069\pm0.005$   & $0.064\pm0.007$   & $0.067\pm0.004$   & $0.061\pm0.005$   & $0.068\pm0.004$     & $0.058\pm0.006$     \\
$N_{CGM}$    & $27\pm4$          & $11\pm1$          & $46\pm4$          & $26\pm4$          & $49\pm4$          & $29\pm4$          & $51\pm5$            & $31\pm4$            \\
$kT_{Coro}$  &                   &                   & $0.70\pm0.03$     & $0.75\pm0.03$     & $0.69\pm0.03$     & $0.71\pm0.03$     & $0.49\pm0.09$       & $0.47\pm0.09$       \\
$\tau$       &                   &                   &                   &                   &                   &                   & $10.7^{+3.5}_{-2.2}$& $9.9^{+3.7}_{-1.8}$\\
$N_{Coro}$   &                   &                   & $0.37\pm0.02$     & $0.36\pm0.02$     & $0.37\pm0.02$     & $0.37\pm0.02$     & $0.9\pm0.2$         & $0.9\pm0.2$         \\
$\chi^2$     & 1129.8            & 1097.9            & 954.5             & 990.3             & 926.8             & 959.8             & 911.0               & 944.6               \\
$dof$        & 815               & 815               & 813               & 813               & 808               & 808               & 807                 & 807                 \\
\hline
\end{tabular}
\end{table*}

\subsection{The signature of the Galactic corona (what is the contribution from faint dwarf M stars?)} 

\begin{figure*}[!th]
\centering
\vspace{-1.0 cm}
\includegraphics[width=0.49\textwidth]{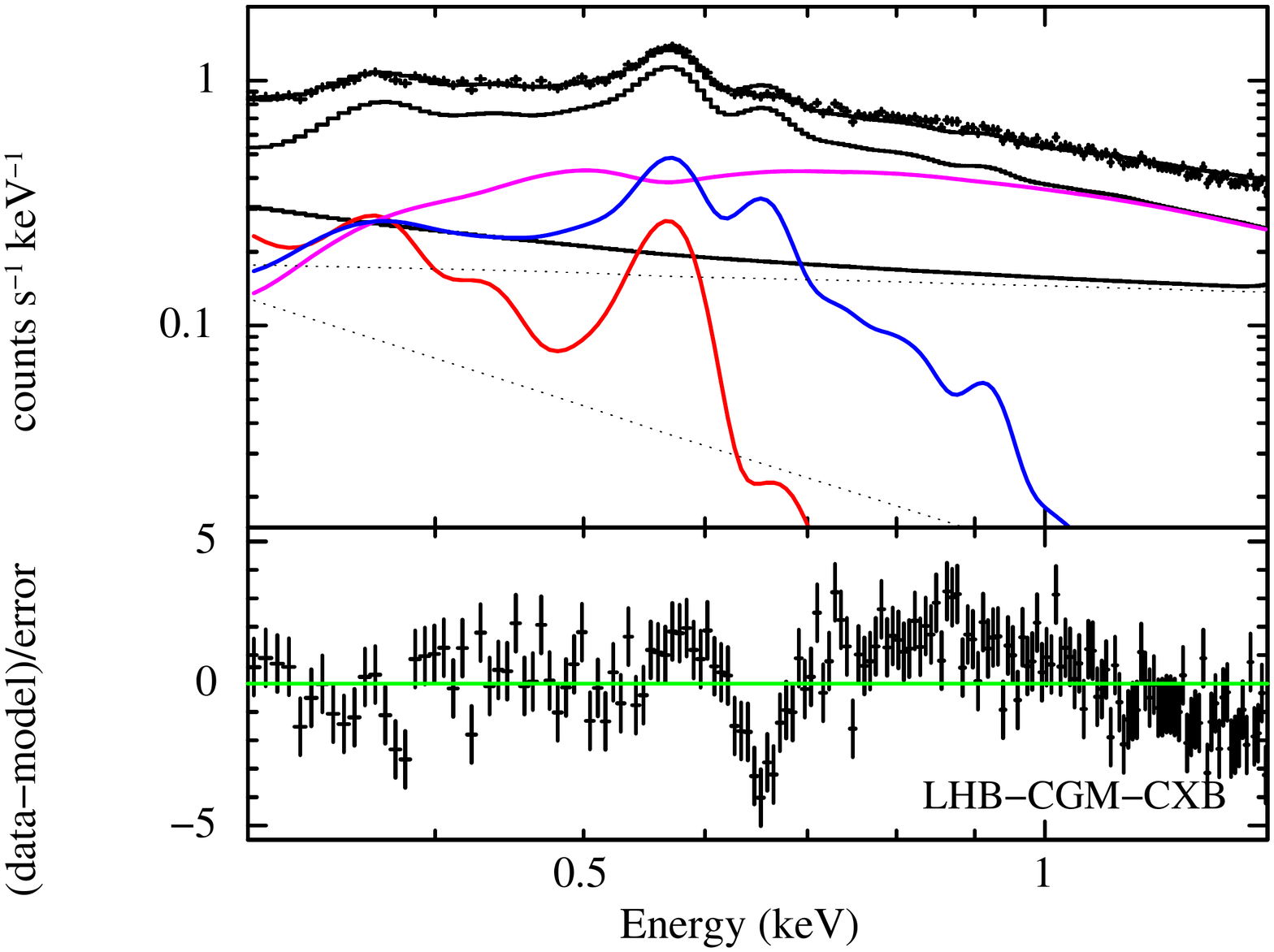}
\includegraphics[width=0.49\textwidth]{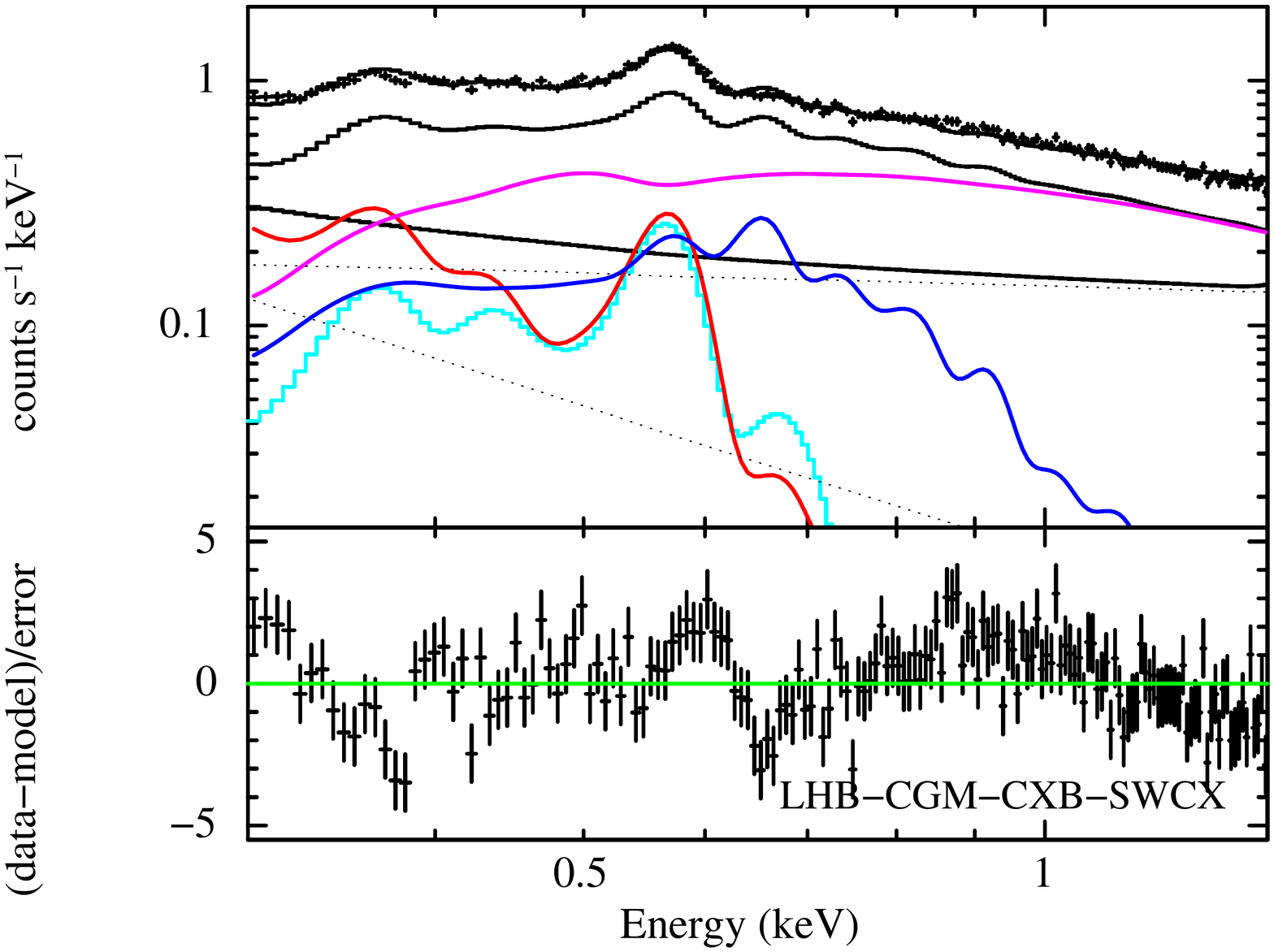}

\vspace{-1.0 cm}
\includegraphics[width=0.49\textwidth]{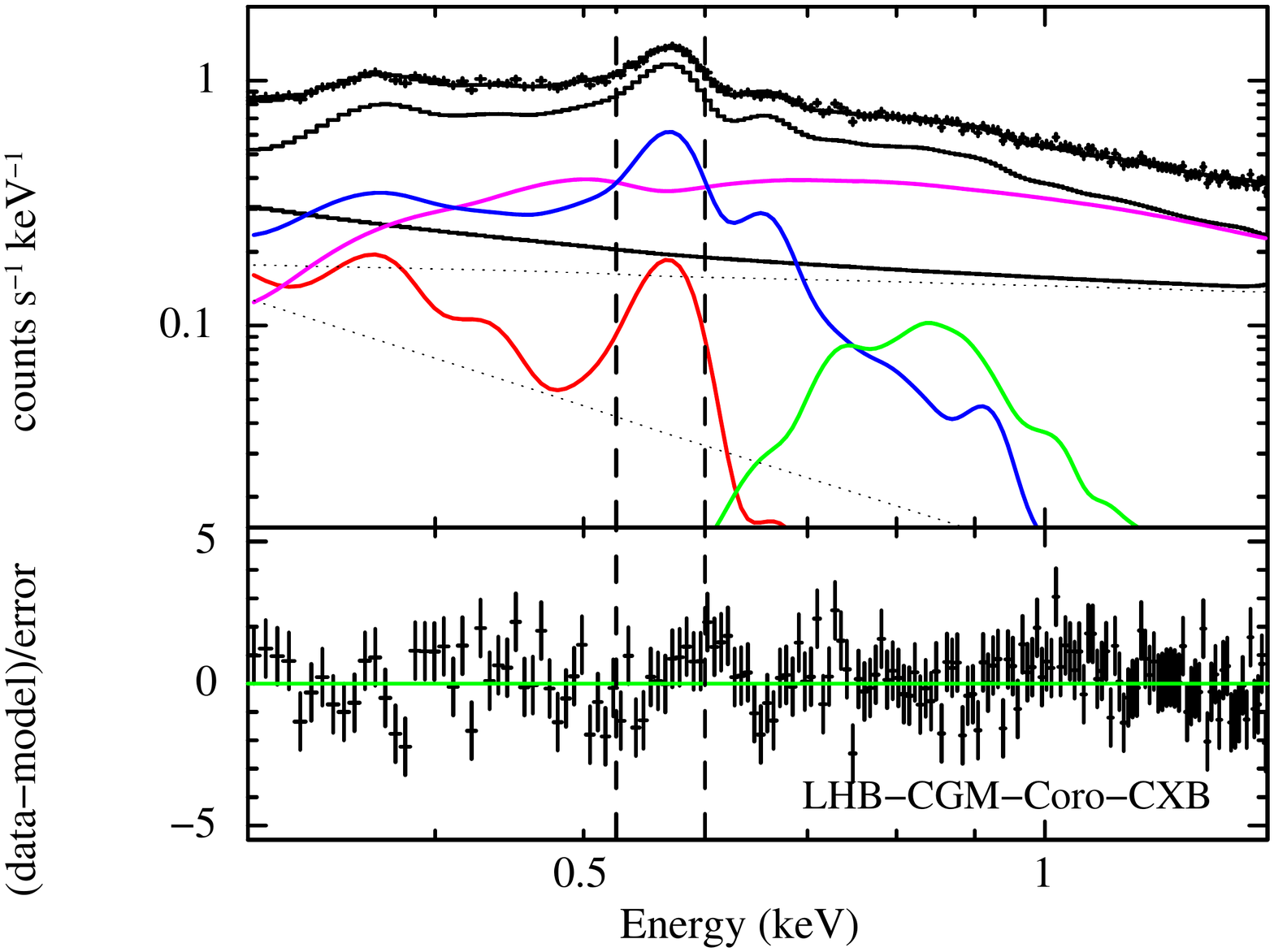}
\includegraphics[width=0.49\textwidth]{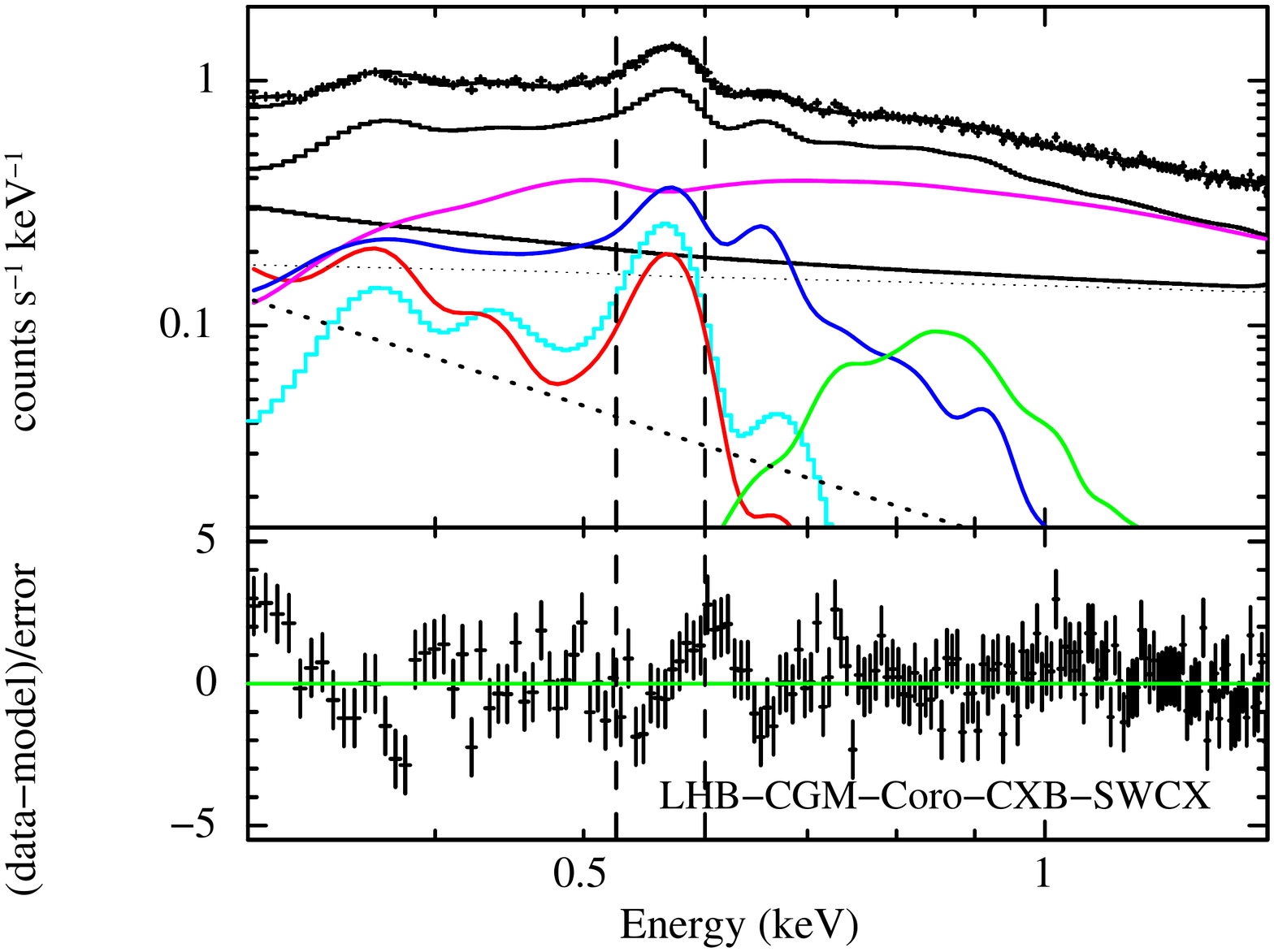}

\vspace{-1.0 cm}
\includegraphics[width=0.49\textwidth]{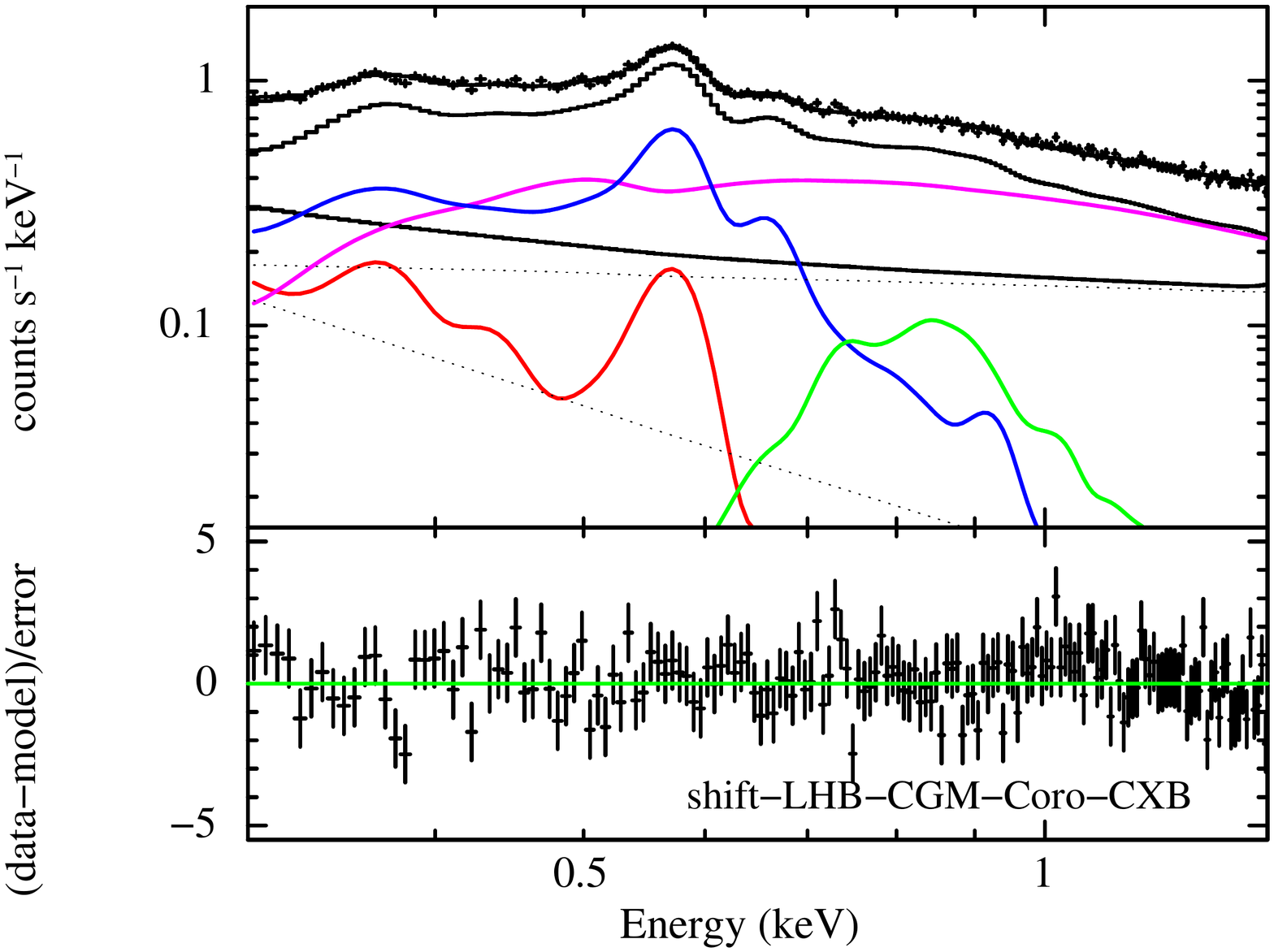}
\includegraphics[width=0.49\textwidth]{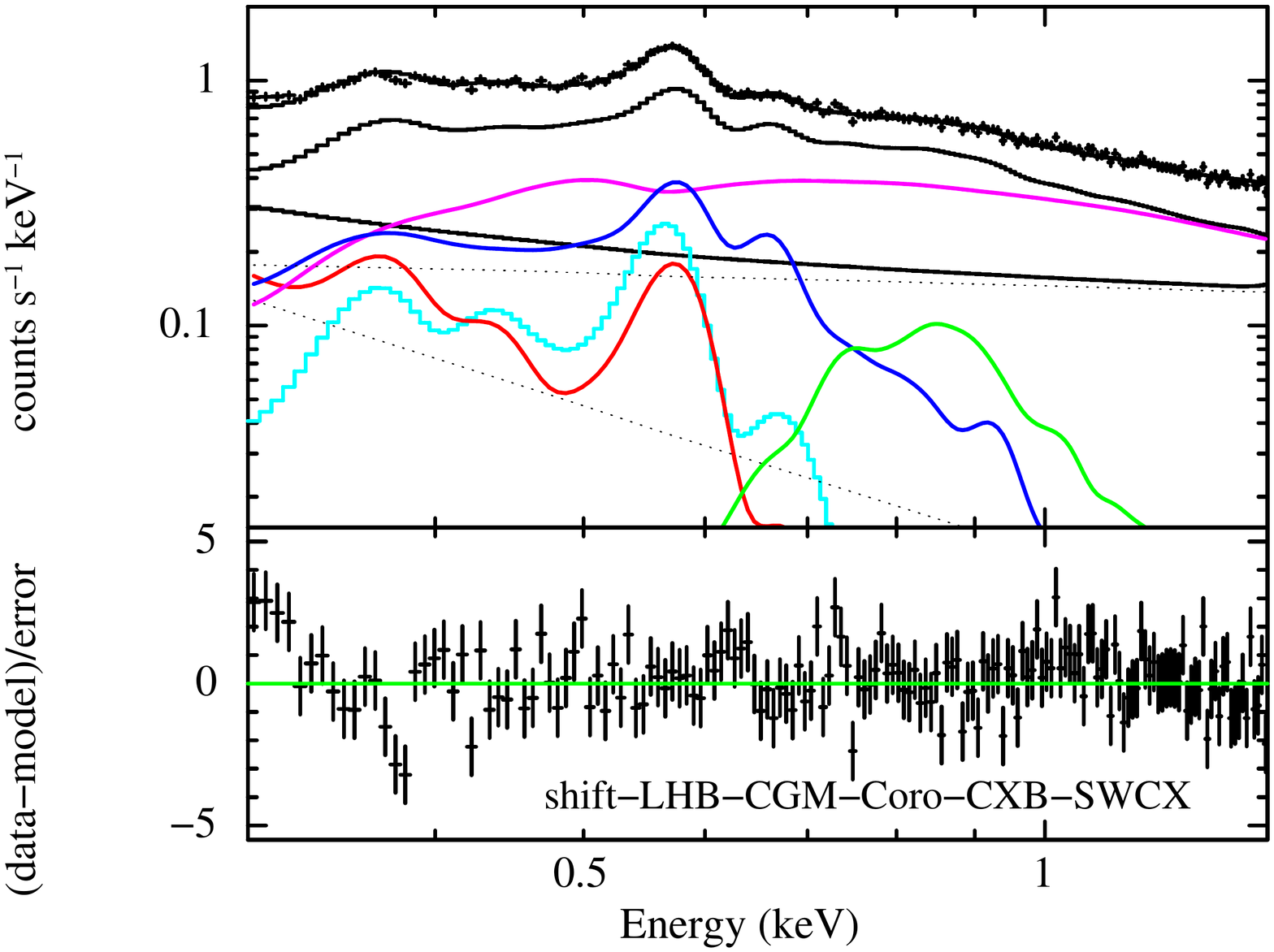}

\vspace{-0.3 cm}
\caption{{\it (Top left panel)} Diffuse emission, as observed by \erosita\ within the \efeds\ field during e12, fitted with a three-component model (LHB-CGM-CXB in Tab. \ref{TCoro}).
The red, blue, magenta and black solid lines show the contribution from the local hot bubble, the circum-Galactic medium, the Cosmic X-ray background and instrumental background, respectively. 
The dotted lines show the various contributions to the instrumental background. 
{\it (Top right panel)} Same as top left panel, once the contribution from SWCX is added to the model (LHB-CGM-CXB-SWCX). 
The cyan line shows the contribution due to SWCX. 
{\it (Central left panel)} The addition of the emission from the Galactic corona (solid green) significantly improves the fit (LHB-CGM-Coro-CXB). However, significant positive residuals (at the position of the vertical dashed lines) remain at the energy of the blue wing of the very prominent \os\ emission line as well as negative residuals on its red wing. 
{\it (Central right panel)} Same as central left panel, with the addition of the SWCX component (LHB-CGM-CXB-Coro-SWCX model).
{\it (Bottom left panel)} The addition of a significantly different energy shift applied to the best fit model of each camera reduces the residuals around the \os\ emission line (shift-LHB-CGM-Coro-CXB). 
{\it (Bottom right panel)} Same as bottom left panel, with the addition of the SWCX component (shift-LHB-CGM-CXB-Coro-SWCX model).
}
\label{FCoro}
\end{figure*}

Regardless of the assumed abundances and whether we include or not the SWCX component, the top panels of Fig. \ref{FCoro} show large residuals at the energy of the \oe\ line and between $\sim0.7$ and $1$ keV. Indeed, the spectrum shows a hump between $\sim0.7$ and $\sim1$~keV which can not be fitted by the power law shape of the CXB (Fig. \ref{FCoro}). 

To reproduce such hump, the continuum of the CGM component would need to have a temperature in excess of $kT\ge0.2$ keV. 
On the other hand, the \os\ / \oe\ line ratio forces the best fit temperature to be lower than $\sim0.2$~keV. 
The result of such "tension" is displayed by the very bad residuals in the top panels of Fig. \ref{FCoro}. 

Such bad residuals represent an incontrovertible evidence that an additional element is required to reproduce the \erosita\ data. 
Therefore, we add to the model a second thermal component to fit the emission from the Galactic corona. 
We initially assume the Galactic corona to be collisionally ionised, to be in thermal equilibrium and optically thin. 
Therefore we assume that it can be described by an {\sc apec} model in {\sc Xspec}. 
We further assume the metal abundance within the Galactic corona to be relatively high (based on the belief that such plasma might be deeply related with fountains and outflows from the interstellar medium in the plane of the Milky Way), therefore we assume solar abundances for this component (Shapiro \& Field 1976; Spitzer 1990; Bregman 1980; Fraternali 2017)\footnote{We performed all fits shown in this paper also assuming a metal abundance of 0.7 solar, obtaining statistically equivalent results. The only parameter in the fit being significantly affected by the change of assumed abundances being the normalisation (emission measure) of the coronal component.}. 

A very significant improvement of the fit is observed by the addition of the spectral component describing the emission of the Galactic corona, which correspond to $\Delta\chi^2=175.3$ and $\Delta\chi^2=107.6$ for the addition of two free parameters in the case of negligible and high SWCX contribution, respectively; see Tab. \ref{TCoro}). 
Indeed, the model comprising the Galactic corona, in addition to the three other components, now reproduces the bulk spectral features in the soft X-ray spectrum (see Fig. \ref{FCoro}). 
The best fit temperature of the Galactic corona is $kT_{Coro}\sim0.70-0.75$ keV, therefore significantly higher than the one of the CGM and somehow smaller but in line with what is typically observed along the Galactic plane (and at the Galactic center) and attributed to the hot phase of the inter-stellar medium (e.g. $kT\sim1$~keV; Ponti et al. 2013 and references therein). 

Along the Galactic disc a thermal component with a temperature of $kT\sim0.7$~keV, therefore with characteristics similar to the Galactic corona, has been observed in \suzaku\ data and it has been attributed to the cumulative emission due to faint dwarf M stars (Masui et al. 2009). 
Additionally, observations with X-ray quantum calorimeters aboard sounding rockets achieved measurements of the soft X-ray emission with unprecedented spectral resolution from four large locations within field of view of $\sim1$ sr (McCammon et al. 2002; 2008; Wulf et al. 2019). 
Wulf et al. (2019) detected a spectral feature at $E\sim0.9$~keV that can be fitted with a hot emission component ($kT\sim0.7$~keV), in addition to the CGM, the LHB and the SWCX emission, in the two fields crossing the Galactic plane, while such emission was absent at high Galactic latitudes. 
Such as Masui et al. (2009), the authors attributed such emission to the contribution due to faint dwarf M stars.  
From the M dwarf model in Wulf et al. 2019, we estimate a contribution from M dwarfs towards the direction of the \efeds\ field to be about half of the observed value of $0.9\times10^{-3}$ pc cm$^{-6}$. 
Therefore, not only stars are expected to give a significant contribution to this coronal emission, but they could, in principle, be dominating over the emission from the Galactic corona, according to the model proposed by Wulf et al. 2019.  

However, we note that the uncertainty on the knowledge of the scale-height of the Galactic disc can induce significant scatter in the predicted contribution due to stars.
Indeed, more recent models of the mass distribution and gravitational potential of the Milky Way refine the disc scale height assumed by Wulf et al. (2019), therefore predicting a different contribution due to stars at the relatively high Galactic latitudes characteristic of the eFEDS field. Despite this will be the subject of a future investigation, carefully addressing this point, we note that the observation of super-virial plasma in absorption towards some bright AGN (Das et al. 2019a; 2021) corroborates the presence of truly diffuse Galactic hot coronal plasma. 

For these reasons, hereinafter we will associate the hot emission towards the \efeds\ field to the Galactic corona, however we stress that a fraction of such emission is, most likely, due to stars. 
Soon, by connecting the improved mass distributions of the Milky Way with the advances in our knowledge of the X-ray emission from stars allowed by \erosita, it will be possible to obtain much improved understanding of the contribution by stars to the overall observed hot plasma emission. 

\subsection{Systematic uncertainty on the energy scale calibration?}
\label{scale}

\begin{figure}[th]
\centering
\includegraphics[width=0.495\textwidth]{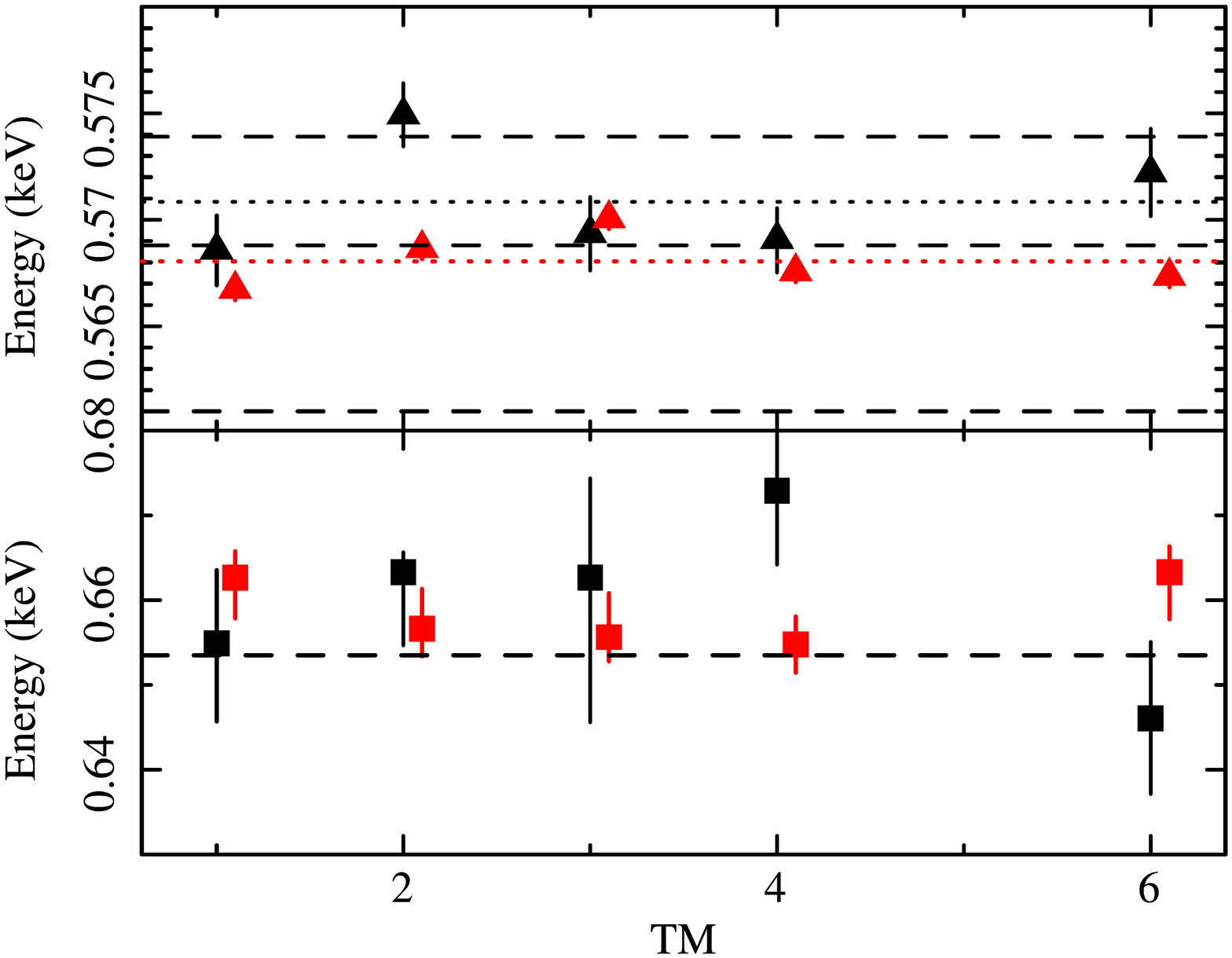}
\caption{Best fit energy of the \os\ and \oe\ emission lines (top and bottom, respectively), as measured by the different cameras aboard \erosita. 
Black and red symbols and the dotted lines show the best fit energies obtained by fitting e12 and e0 spectra of each TM camera and the average best fit energy, respectively.
A scatter larger than the statistical uncertainties is observed during both e0 ($\Delta E\sim3$~eV) and e12 ($\Delta E\sim5$~eV). 
The horizontal dashed lines indicate the expected energy of the transitions (three for the \os\ Helium like triplet). }
\label{Flines}
\end{figure}
The middle panels of Fig. \ref{FCoro} still show positive residuals at the energies of the blue wing of the \os\ emission line and negative residuals at the energy of the red wing of the same line. 

To investigate the origin of such residuals, we fitted the spectrum with a parametric continuum model plus an array of emission lines.
To perform this task, we further restricted the energy band over which we perform the fit to the $0.3-0.9$~keV energy range, where the most prominent emission lines dominate over the continuum. 
We fit the continuum with a power law (with photon index free to vary) plus a thermal component with no emission line ({\sc apec} component with no emission line). 
Additionally, we add four emission lines, to reproduce the \os, the \oe\ and the \cs\ emission, plus a weaker line at $E\sim0.423$~keV, which reproduces the N {\sc vi} triplet. 

The energies and normalisations of the emission lines are free to vary, while their widths are fixed to $\sigma=1$~eV for the Hydrogen like lines, while to $\sigma=4$~eV for Helium like lines, to account for the ensemble of the triplet lines. 
The top and bottom panels of Fig. \ref{Flines} show the best fit energy of the \os\ and \oe\ emission lines, respectively, as observed by the different cameras aboard \erosita, during e12 and e0 in black and red, respectively. 
The horizontal dashed lines indicate the expected energies of the lines (the three lines show the energies of the \os\ triplet). 
The error bars reported in Fig. \ref{Flines} correspond to the $1 \sigma$ statistical uncertainty, as derived from the fit. 

The energy of the \os\ line in the e0 spectra (which has the highest signal to noise) is determined with high precision (see Fig. \ref{Flines}). 
Such small statistical uncertainties allow us to reveal the systematic uncertainties associated with the reconstructed absolute line energy for each camera, which does not exactly match the incident line energy (see Dennerl et al. 2020). 
In particular, we observe a systematic shift which can be as large as $\Delta E \sim2-3$ eV at the energy of the \os\ line\footnote{The larger statistical uncertainties prevent us from investigating the systematic scatter in either the \oe\ line or the \os\ line in the lower statistic e12 spectrum. }, which is consistent with the current calibration of the energy scale (Dennerl et al. 2020). 

From the results above, we conclude that the analysis of spectra containing emission lines is very demanding with respect to the energy calibration, as errors in the absolute energy scale by a few eV can already cause significant residuals and may lead to wrong conclusions. 
In order to mitigate this problem, we allow for fine adjustments of the absolute energy scale in the spectral fits of the total spectrum by using a velocity shift ({\sc vashift} component in Xspec; see Tab. \ref{TCoro} and bottom panel of Fig. \ref{FCoro}). 
Although this technique should conceptually be understood only as an approximation, the resulting energy shifts seem to be sufficiently small to justify this simplified approach. 
Indeed, a significant improvement of the fit ($\Delta\chi^2=27.7$ and $\Delta\chi^2=30.5$ for the addition of 5 free parameters in the case of negligible and high SWCX contribution, respectively; see Tab. \ref{TCoro}) is observed by the addition of a shift of the energy scale. 
The standard deviation of the observed shifts is $\sigma\sim1$~eV, at the \os\ line energy, and it can be as large as $\sim3$~eV. 

\subsection{Is the Galactic corona in thermal equilibrium?} 

\begin{figure*}[th]
\centering
\vspace{-1.0 cm}
\includegraphics[width=0.49\textwidth]{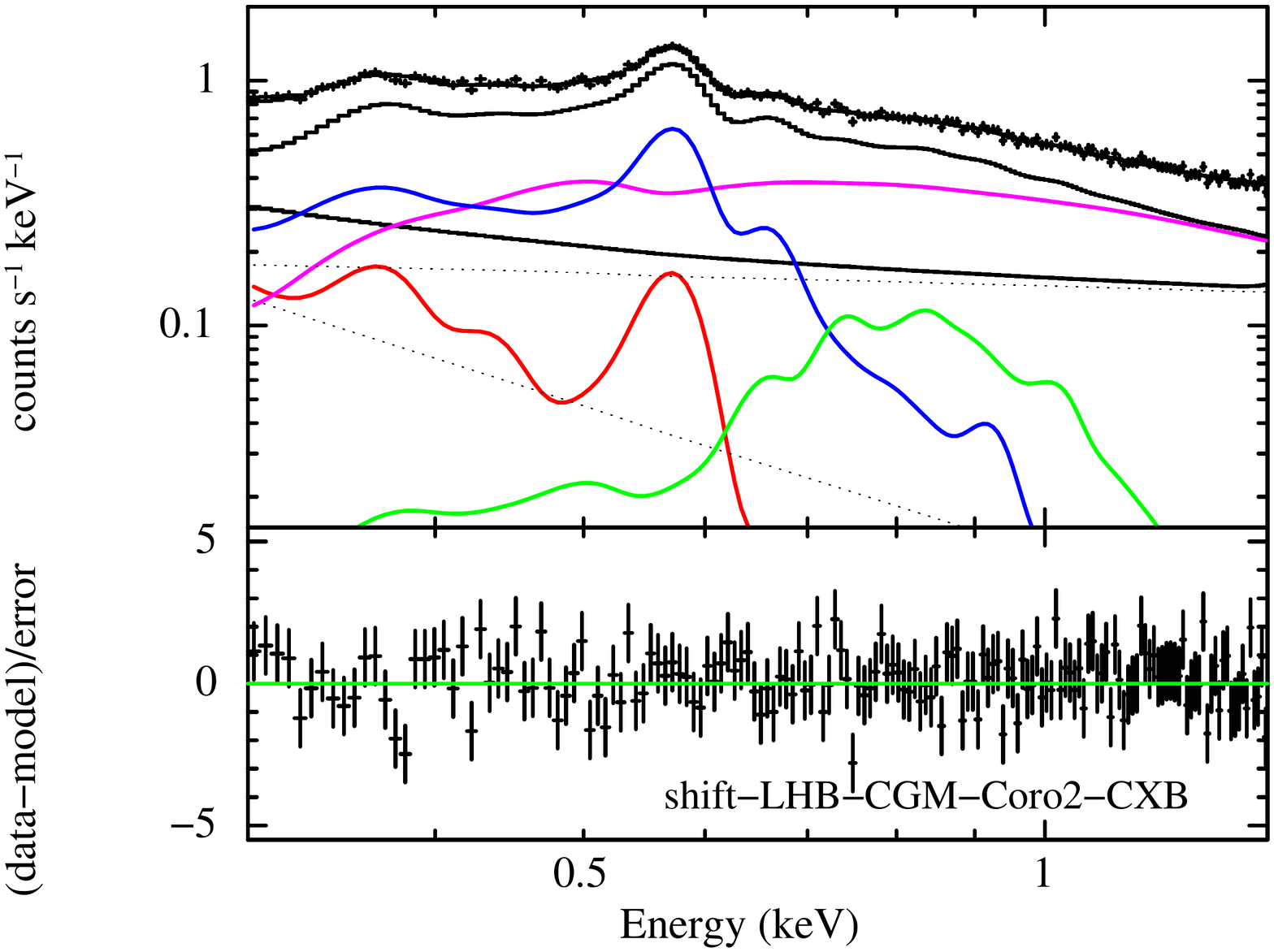}
\includegraphics[width=0.49\textwidth]{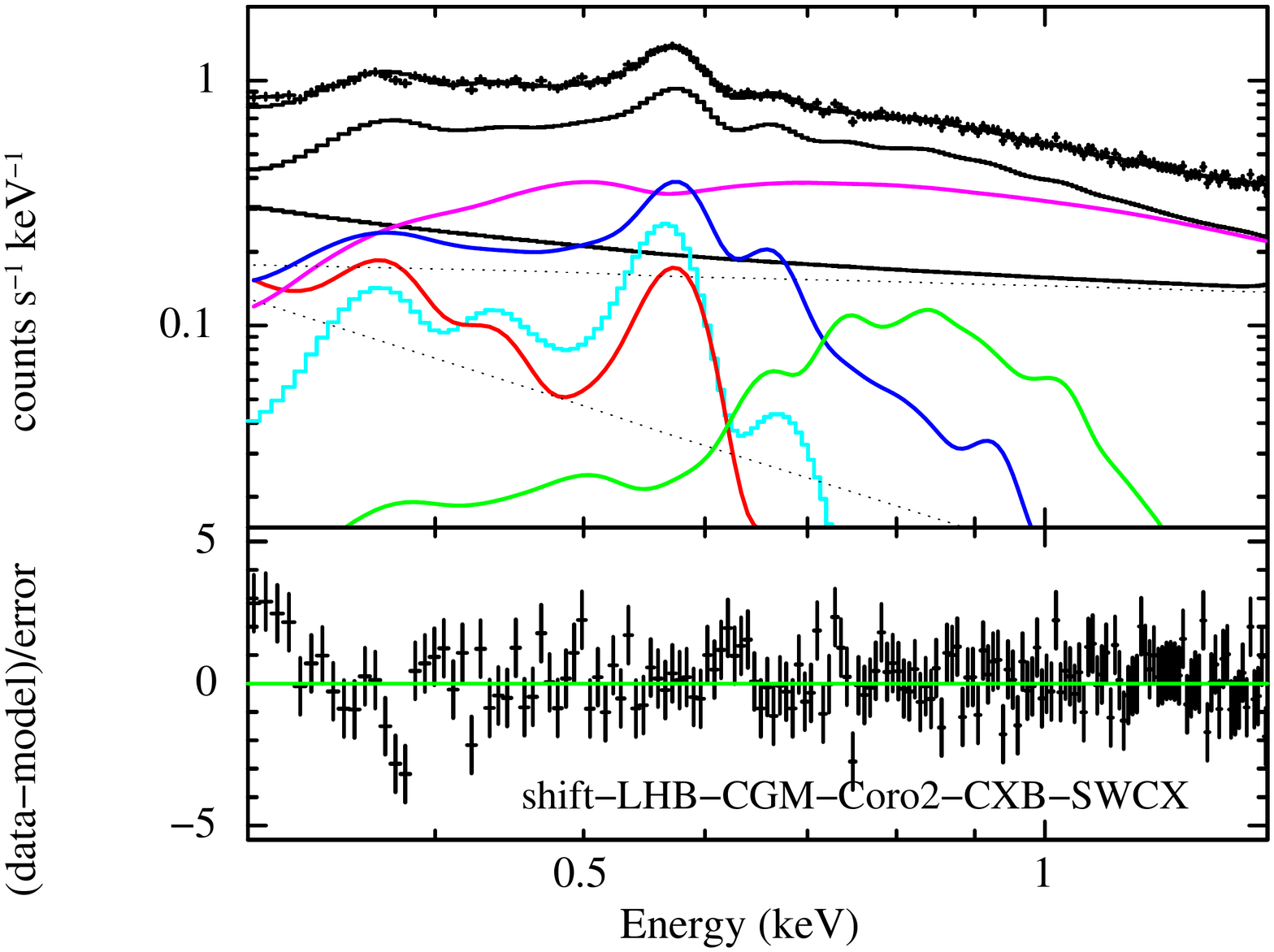}

\vspace{-0.3 cm}
\caption{Same spectrum and color scheme as in Fig. \ref{FCoro}. 
{\it (Left panel)} Best fit result with a recombining plasma emission model ({\sc rnei} in {\sc Xspec}) for the Galactic corona (solid green) in addition to the components considered in the bottom left panel of Fig. \ref{FCoro} (shift-LHB-CGM-Coro2-CXB in Tab. \ref{TCoro}).
{\it (Right panel)} Same as left panel, with the addition of the SWCX component (shift-LHB-CGM-CXB-Coro2-SWCX model).
}
\label{FCoro2}
\end{figure*}
To investigate the thermal equilibrium of the plasma in the Galactic corona we substitute the {\sc apec} model for the Galactic corona with a recombining plasma model ({\sc rnei} model in {\sc Xspec}). Such model is reproducing the emission from plasma which was initially hot and then rapidly cooled on a time-scale shorter than the one required to reach thermal equilibrium. 
Indeed, as a consequence of the low densities characteristic of the Galactic corona, we might expect that such plasma might be out of thermal equilibrium. 
One possible scenario for this might be thought in the form of a hot outflow (or the rising part of a fountain) from the Galactic disc which maintains the Galactic corona constantly replenishing it with: energy; plasma; metals; energetic particles; etc. (Bregman et al. 1980; Fraternali et al. 2015; Putman et al. 2012). 
In addition to solar abundances, we also assumed an initial plasma temperature of $1.2$~keV, to match the typical temperatures of the hot plasma in the Galactic disc, 
while we left the current plasma temperature ($kT_{Coro}$) as a free parameter in the fits. 

A significant improvement of the fit ($\Delta\chi^2=15.8$ and $\Delta\chi^2=15.2$ for the addition of 1 free parameter in the
case of negligible and high SWCX contribution, respectively; F-Test probability of $3\times10^{-4}$; see Tab. \ref{TCoro}) is observed once a recombining plasma model is used. 
Indeed, both panels of Fig. \ref{FCoro2} show that the recombining plasma component is able to better reproduce the data leaving lower residuals in the $\sim1$~keV band. 

The best fit plasma temperature are $kT_{Coro}=0.49\pm0.09$~keV and $kT_{Coro}=0.47\pm0.09$~keV in the case of negligible and high SWCX contribution, respectively, therefore significantly higher than the one of the circum-Galactic medium. 
From the normalisation of the coronal emission, we derive an emission measure of $0.9\pm0.3\times10^{-6}$ cm$^{-6}$ kpc, which would correspond to an electron density of $n_e\sim0.9\times10^{-3}$ cm$^{-3}$ for a depth of $\sim1$~kpc. 
Assuming that the recombining plasma model is an accurate description of the coronal emission, the best fit provides us with an estimate of the ionisation time-scale which results to be $\tau=(10.7\pm3)$ and $(9.9\pm3)\times10^{10}$~s~cm$^{-3}$ in the
case of negligible and high SWCX contribution, respectively (see Tab. \ref{TCoro}). 
For an electron density of $n_e\sim0.9\times10^{-3}$~cm$^{-3}$, this time-scale would correspond to an ionisation time-scale of about $\sim4$~Myr, which is longer than the time-scale needed for an outflow originating from hot plasma ($kT\sim1$~keV) in the Galactic disc and moving at the sound speed ($v_c\sim500$~km s$^{-1}$) in an outflow replenishing the Galactic corona. Indeed, such hot plasma would be able to cover $\sim1$~kpc in $\sim2$ Myr, which seems in line with the observation that the coronal plasma is possibly out of thermal equilibrium. 

Both panels of Fig. \ref{FCoro2} show that, even at its peak, the emission from the Galactic corona is comparable, but lower, than the instrumental background and a factor of $\sim2.5-3$ times lower than the emission from the CXB.
The relative weakness of the emission from the Galactic corona is in line with the fact that it has been recognised as a separate feature, in addition to the Galactic halo component and different from the emission from dwarf M stars (Masui et al. 2009; Wulf et al. 2019), only recently (Das et al. 2019a,b; 2021). 

\section{Constraining the properties of the hot CGM} 
\label{SectCGM}

The \erosita\ data allow us to place robust constraints on the physical properties of the CGM. 

\subsection{Temperature of the CGM constrained by the \os\ and \oe\ emission lines}
\label{lines}

For optically thin hot plasma in thermal equilibrium, the energy and intensity of the emission lines can be used as a powerful tool to estimate the temperature of the plasma, independently from the shape of the underlying continuum.

During e12, the best fit energy of the \os\ line is $E_{\rm O {\sc vii}}\sim0.571$ keV (black dotted line in Fig. \ref{Flines}), therefore consistent with being dominated by the recombination line, as expected in the case of collisionally ionised plasma. 
On the other hand, the best fit energy of the \os\ line is observed to shift to $E_{\rm O {\sc vii}}\sim0.568$~keV during e0 (black dotted line in Fig. \ref{Flines}). 
We attribute such shift of the best fit energy to a larger contribution of the forbidden line, which is dominant in the SWCX component. 
Therefore, this corroborates the idea that the higher flux of the \os\ emission line during e0 is produced by enhanced SWCX. 

For the \os\ line in the e0 spectrum, such statistical uncertainties are significantly smaller than the differences in energies ($\Delta E\sim2-3$~eV) measured by the different instruments, therefore confirming that the observed scatter is due to systematic uncertainties in the calibration of the energy scale of the different cameras aboard \erosita. 

To measure the \os\ G-ratio\footnote{The G-ratio is defined as: G=(f+i)/r, where f, i and r are the intensities of the forbidden, intercombination and recombination lines, respectively (Porquet \& Dubau 2000).} and the \os\ over \oe\ line ratio, we fit the spectra of each TM with the same parametric model used in Sect. 7.2 (composed of a power law plus an apec component with no emission line), however we substitute the four emission lines with six narrow emission lines with Gaussian profiles (three for the \os\ triplet, plus \cs, \oe\ and the weaker line at $\sim0.423$~keV, which reproduces the N {\sc vi} triplet) with energies fixed at the expected values of each transition and shifted by a common value for each camera (produced by the {\sc vashift} component in {\sc Xspec}). 
To avoid degeneracy due to the fact that the lines of the \os\ triplet cannot be separated at the CCD resolution of the \erosita\ cameras (Predehl et al. 2021), we fix the ratio of the forbidden over intercombination lines at 3.5, as expected for plasma at densities as low as the ones considered here. 
We expect that the combination of the \cs\ and \oe\ lines will constrain the cross-calibration across the different cameras by determining the inter-camera shift of the energy scale. 
This will then allow us to estimate the G-ratio of the \os\ triplet (Porquet et al. 2000; 2001). 

\begin{figure}[t]
\vspace{-0.7cm}
\centering
\includegraphics[width=0.495\textwidth]{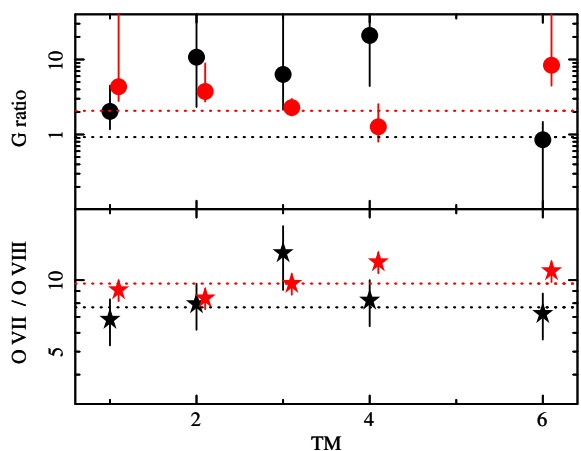}
\caption{Best fit G ratio and \os\ / \oe\ emission lines ratio as observed by the different cameras in the top and bottom panels, respectively. Same color scheme as in Fig. \ref{Flines}}
\label{Flines2}
\end{figure}

\begin{figure}[th]
\centering
\includegraphics[width=0.48\textwidth]{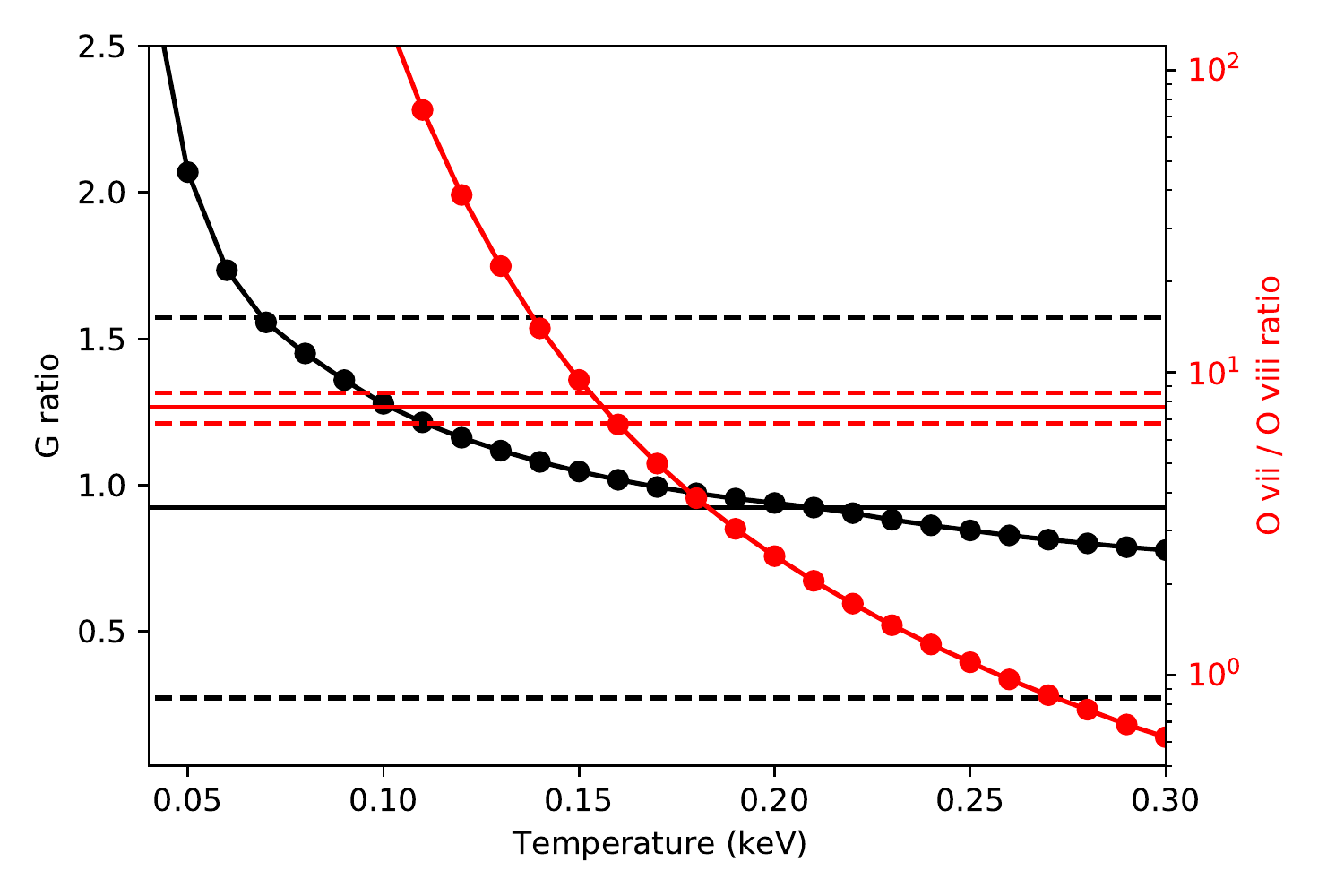}
\caption{The black solid and dashed horizontal line shows the best fit and 1~$\sigma$ uncertainty on the G-ratio measured by fitting the lines of the \os\ triplet from the e12 spectrum. 
The black filled dots and connecting line show the expected relation between the G-ratio and temperature, expected for collisionally ionised plasma in thermal equilibrium (obtained from the {\sc apec} model in Xspec). 
The measurement of the G-ratio implies that the temperature must be $kT>0.06$~keV, if produced by a single plasma component in thermal equilibrium. 
The red solid and dashed horizontal line shows the best fit and 1~$\sigma$ uncertainty on the \oe\ over \os\ line ratio measured by fitting the e12 spectrum. 
The red filled dots and connecting line show the expected relation between \oe\ over \os\ line ratio and temperature, expected for collisionally ionised plasma in thermal equilibrium (obtained from the {\sc apec} model in Xspec). 
Under the assumption that a single optically thin and collisionally ionised component is producing both lines, the observed line ratio implies that the temperature must be within the range $0.152<kT<0.160$~keV. 
The two completely independent estimates of the temperature are in agreement with each other, suggesting that the same plasma might produce the bulk of the \os\ and \oe\ emission. }
\label{Flines3}
\end{figure}

The black and red points in the top panel of Fig. \ref{Flines2} show the G-ratio of the \os\ triplet in the e12 and e0 spectra, respectively\footnote{The ability to determine the G-ratio through this technique is somehow hindered by leaving the energy scale anchored to the \cs\ and \oe\ lines, which are significantly weaker than the \os\ line. This is then reflected into the rather large error bars associated with the determination of the G-ratio. }. 
Because of the large error bars and scatter associated with the determination of the G-ratio, the top panel of Fig. \ref{Flines2} shows the y-axis in logarithmic scale. 
G-ratios as large as 20 are observed, however these large values are characterised by very large uncertainties (Fig. \ref{Flines2}). 
We compute the best fit G-ratio by performing a fit with a constant, which is equivalent to a weighted mean. 
In particular, we observe that during e12 a fit of the G-ratio observed by each camera provides the best fit value of $G=0.9^{+0.6}_{-0.5}$. 
It is well known that the G-ratio is a good temperature diagnostic (Porquet et al. 2000; 2001). 
Indeed, by comparing the measured G-value and its uncertainties (dashed lines in Fig. \ref{Flines3}) with the expectations from the models of collisionally ionised plasma (black circles in Fig. \ref{Flines3}), we constrain that the temperature of the plasma producing the \os\ triplet must be larger than $kT>0.06$~keV. 

The bottom panel of Fig. \ref{Flines2} shows the ratio of the best fit intensities of the \os\ and \oe\ lines, obtained by fitting the total spectrum. 
The fit of the ratio with a constant provides a best fit value of about $7.7\pm0.9$. 
Such lines ratio is also a very sensitive temperature diagnostic tool. 
Indeed, the red circles in Fig. \ref{Flines3} show the expected ratio as a function of the temperature of the plasma\footnote{Such lines show the G-ratio and line ratios as expected for optically thin collisionally ionised plasma in thermal equilibrium (the {\sc apec} model in Xspec). }, under the assumption that a single optically thin and collisionally ionised component is producing all of the flux from the \os\ and \oe\ lines.  
By comparing our measurement with the expected relation (see the dashed red lines in Fig. \ref{Flines3}), we place a tighter constraint on the temperature of the plasma producing the \os\ and \oe\ lines of $0.152<kT<0.160$~keV\footnote{As for the G-ratio, the larger \os\ over \oe\ ratio observed during e0 is a consequence of the additional contribution due to SWCX. }. 

The best fit temperature of the CGM component results to be $kT=0.157\pm0.004$~keV, while it increases to $kT=0.173\pm0.005$~keV for a high SWCX contribution. Indeed, if the SWCX component provides a significant fraction of the \os\ line, then the temperature estimate derived from the line ratio will be biassed low. 

We conclude that, despite the statistical uncertainty on the measurement of the CGM temperature is as low as $\sim3$~\%, the systematic uncertainty induced by the uncertainty on the amplitude of the SWCX emission during e12 is as large as $\sim10$~\%. In particular, assuming a larger contribution due to SWCX does result in a hotter CGM plasma (i.e. $kT=0.173\pm0.005$~keV). 

Finally, we estimate an upper limit to the soft X-ray line widths. 
Indeed, after assuming that all 6 lines, which have been fitted, are broadened by the same amount, we measured an upper limit to the line widths of $\Delta v \leq500$~km s$^{-1}$. 

\subsection{Determining the metal abundances of the hot CGM}
\label{CGMAb}

We note that the best fit metal abundances of the CGM component is $Z_{CGM}=0.068\pm0.004$~$Z_{\odot}$ and $Z_{CGM}=0.058\pm0.006$~$Z_{\odot}$ for negligible and high SWCX contribution, respectively, which correspond to a statistical uncertainty of the order of $\sim5$~\%. 
Such remarkable statistical accuracy is due to the fact that the CGM component produces both the soft X-ray emission lines as well as most of their underlying continuum. 
The systematic uncertainty on the metallicity induced by the poorly constrained SWCX contribution is of the order of only $\sim15$~\%. 
In particular, the best fit with high SWCX contribution corresponds to lower abundances, in agreement with the fact that SWCX provides a larger contribution to the lines than to the continuum. 

The best fit metal abundance appears to be remarkably low, with best fit values of the order of $Z_{CGM}\sim0.06-0.07$~$Z_{\odot}$, for the Lodders (2003) abundances, regardless of the model employed (see Tab. \ref{TCoro}). 
This is dictated by the low equivalent widths of the soft X-ray emission lines, such as \os\ and \oe. 
Indeed, both Fig. \ref{FCoro} and \ref{FCoro2} show that the thermal component associated with the emission of the CGM reproduces not only the bulk of the emission lines, but also the bulk of the soft X-ray continuum between such lines. 
In the next sections, we will investigate the order of magnitude of systematic uncertainties on the estimated abundances. 

\subsubsection{The impact of the CXB on the measured CGM abundances}
\label{CXBs}

As detailed in Sect. \ref{CXB}, the shape of the CXB is well known above $\sim1$~keV, while significant uncertainties are related with its contribution below $\sim1$~keV. To take into account the impact of such uncertainties on the determination of the metal abundances of the CGM component, we re-fitted the \efeds\ spectrum substituting the CXB component first with its harder possible spectrum (CXBh; see \S~\ref{CXB}). 

\begin{table*}
\small
\centering
    \caption{Best fit parameters obtained by fitting the e12 spectrum with different models.
    Same nomenclature as for Tab. \ref{TCoro}. 
    CXBh and CXBs mean that the CXB component has been substituted with the CXBh and CXBs ones. 
    CXB-PL considers the CXB component, with the addition of an absorbed power law at low energy (see \S~\ref{AddPL}). 
    $\Gamma_{CXBPL}$ and $N_{CXBPL}$ report the photon index and normalisation (in units of photons keV$^{-1}$ cm$^{-2}$ s$^{-1}$ at 1 keV) of such additional non-thermal diffuse component.
    $\dag$ Value fixed in the fit.}
    \label{CGMAbun}
    \begin{tabular}{c | c c | c c | c c }
    \hline \hline
    \multicolumn{7}{c}{\bf SPECTRUM e12} \\
    \hline \hline
             & \multicolumn{2}{c}{shift}               & \multicolumn{2}{c}{shift}                & \multicolumn{2}{c}{shift}                 \\
             & \multicolumn{2}{c}{LHB-CGM-Coro2-CXBh}  & \multicolumn{2}{c}{LHB-CGM-Coro2-CXBs}   & \multicolumn{2}{c}{LHB-CGM-Coro2-CXB-PL}    \\
             &                    &  SWCX              &                     & SWCX               &                           & SWCX            \\
$N_{LHB}$    & $2.7\pm0.5$        & $3.1\pm0.5$        & $2.7\dag$           & $2.7\dag$           & $2.7\dag$                & $2.7\dag$       \\
$N_{CXB}$    & $0.343\pm0.004$    & $0.341\pm0.004$    & $0.327\pm0.004$     & $0.325\pm0.004$     & $0.264\pm0.005$          & $0.267\pm0.004$ \\
$\Gamma_{CXBPL}$&                 &                    &                     &                     & $4.3\pm0.3$              & $4.8\pm0.4$         \\
$N_{CXBPL}$     &                 &                    &                     &                     & $0.020\pm0.006$          & $0.009\pm0.004$ \\
$kT_{CGM}$   & $0.157\pm0.004$    & $0.171\pm0.006$    & $0.153\pm0.004$     & $0.167\pm0.007$     & $0.156\pm0.003$          & $0.172\pm0.004$ \\
$Z_{CGM}$    & $0.057\pm0.003$    & $0.046\pm0.003$    & $0.093\pm0.008$     & $0.089\pm0.009$     & $0.3\dag$                & $0.3\dag$   \\
$0.57\pm0.03$\\
$N_{CGM}$    & $65\pm5$           & $44\pm4$           & $40\pm2$            & $22\pm2$            & $12.5\pm0.5$             & $7.1\pm0.4$\\
$kT_{Coro}$  & $0.47\pm0.08$      & $0.45\pm0.08$      & $0.49\pm0.11$       & $0.50\pm0.10$       & $0.49\pm0.09$            & $0.47\pm0.09$  \\
$\tau$       &$10.2^{+2.9}_{-1.8}$&$9.5^{+2.9}_{-1.4}$ & $10.5^{+4.5}_{-2.5}$& $10.6^{+4.5}_{-2.3}$& $11.1^{+4.1}_{-2.5}$     & $10.2^{+3.6}_{-1.9}$ \\
$0.07^{+0.03}_{-0.02}$\\
$N_{Coro}$   & $1.0\pm0.2$        & $1.1\pm0.2$        & $0.8\pm0.2$         & $0.8\pm0.2$         & $0.8^{+0.4}_{-0.1}$      & $0.9^{+0.4}_{-0.2}$\\
$\chi^2$     & 913.1              & 948.8              & 923.7               & 964.3               & 908.3                    & 933.8                \\
$dof$        & 807                & 807                & 808                 & 808                 & 807                      & 807       \\
\hline
\end{tabular}
\end{table*}
The left columns of Tab. \ref{CGMAbun} show the best fit results once the CXB component is substituted with the CXBh one, for both the case of negligible or high SWCX flux in the first and second column, respectively (Tab. \ref{CGMAbun}).
We observe that, as a result of the introduction of the CXBh component, the best fit CGM abundance drops significantly to values of $Z_{CGM}=0.057\pm0.003$ and $Z_{CGM}=0.046\pm0.003$~$Z_{\odot}$ and the quality of the fit worsen ($\Delta\chi^2=-2.1$ and $-4.2$ for the same degrees of freedom) in the case of negligible and high SWCX scenarios, respectively. 
This confirms the expectation that, if a smaller fraction of the soft X-ray continuum is produced by the CXB, then the CGM will be required to have smaller metal abundances to produce the same lines and vice-versa. 

We then re-fitted the \efeds\ spectrum substituting the CXB component with its softer possible spectrum (CXBs; see \S~\ref{CXB}). 
Once the spectrum if fitted with this model, we observe that the normalisation of the LHB component rises to high values ($N_{LHB}=0.0041\pm0.0004$~pc~cm$^{-6}$), which are inconsistent with the ones observed by \rosat\ ($N_{LHB}\sim0.0027$~pc~cm$^{-6}$). 
Indeed, would the LHB emission be so high, then it would predict a flux in the R1 and R2 \rosat\ bands significantly larger than the observed ones. 
Therefore, we fix the normalisation of the LHB to the value observed by \rosat. 

The central columns of Tab. \ref{CGMAbun} show the best fit results, once the CXB component is substituted with the CXBs one, for both the case of negligible or high SWCX flux, respectively (Tab. \ref{CGMAbun}).
Also this time the quality of the fit worsen ($\Delta\chi^2=-12.7$ and $-19.8$ for one less degree of freedom), however we observe that the best fit CGM abundance rises significantly to values of $Z_{CGM}=0.093\pm0.008$ and $Z_{CGM}=0.089\pm0.009$~$Z_{\odot}$, in the case of negligible and high SWCX, respectively (Tab. \ref{CGMAbun}). 
This suggests that CGM abundances as high as $Z_{CGM}=0.1$~$Z_{\odot}$ are not excluded by the data, if a soft CXB component is assumed. 
In particular, this exercise shows that, despite the statistical uncertainty on the measurement of the CGM abundance is as small as $\sim5$~\%, the uncertainty on the true contribution to the soft X-ray continuum of the CXB component induces a larger systematic uncertainty of the order of $\sim70$~\%. 
In fact, changing the assumptions on the shape of the CXB below $\sim1$ keV, we measure an abundance within the range $Z=0.046\pm0.003$ to $Z=0.093\pm0.008$~$Z_{\odot}$.

\subsubsection{An additional non-thermal component to the soft X-ray diffuse emission?}
\label{AddPL}

An alternative option to recover larger metal abundances for the CGM would be to assume that a new hypothetical component would produce the bulk of the continuum in the $\sim0.3-1$~keV band. 
Such component shall not produce emission lines, therefore it must be non-thermal\footnote{We deem as unlikely the possibility that a thermal and optically thick component, such as black body emission, could be associated with the rarefied plasma producing the diffuse emission over large portions of the sky in extra-galactic fields. }. 
Additionally, such hypothetical component must be truly diffuse and be relevant only at low energies, providing a contribution smaller than $\sim10$ \% of the CXB at energies above $\sim0.5$~keV. 
Indeed, ultra-deep X-ray surveys with \chandra\ and \xmm\ have resolved more than $\sim92$~\% of the CXB in the 0.5-7 keV band (Luo et al. 2017). 

To investigate such possibility, we add a power law component to the fit, which we assume to be absorbed by the full column density of absorbing Galactic material. 
Then, we fix the metal abundance of the CGM component to the significantly larger value of $Z_{CGM}=0.3$ $Z_\odot$. 
Finally, we constrain the normalisation of the LHB to be consistent with the value observed by \rosat\ and the normalisation of the CXB to lay within 10 per cent of its expected value ($N_{CXB}=0.269$ ph keV$^{-1}$ cm$^{-2}$ s$^{-1}$ at 1 keV), therefore it is constrained to lay within 0.242-0.296 ph keV$^{-1}$ cm$^{-2}$ s$^{-1}$ at 1 keV. 
\begin{figure*}[t]
\centering
\includegraphics[width=0.49\textwidth]{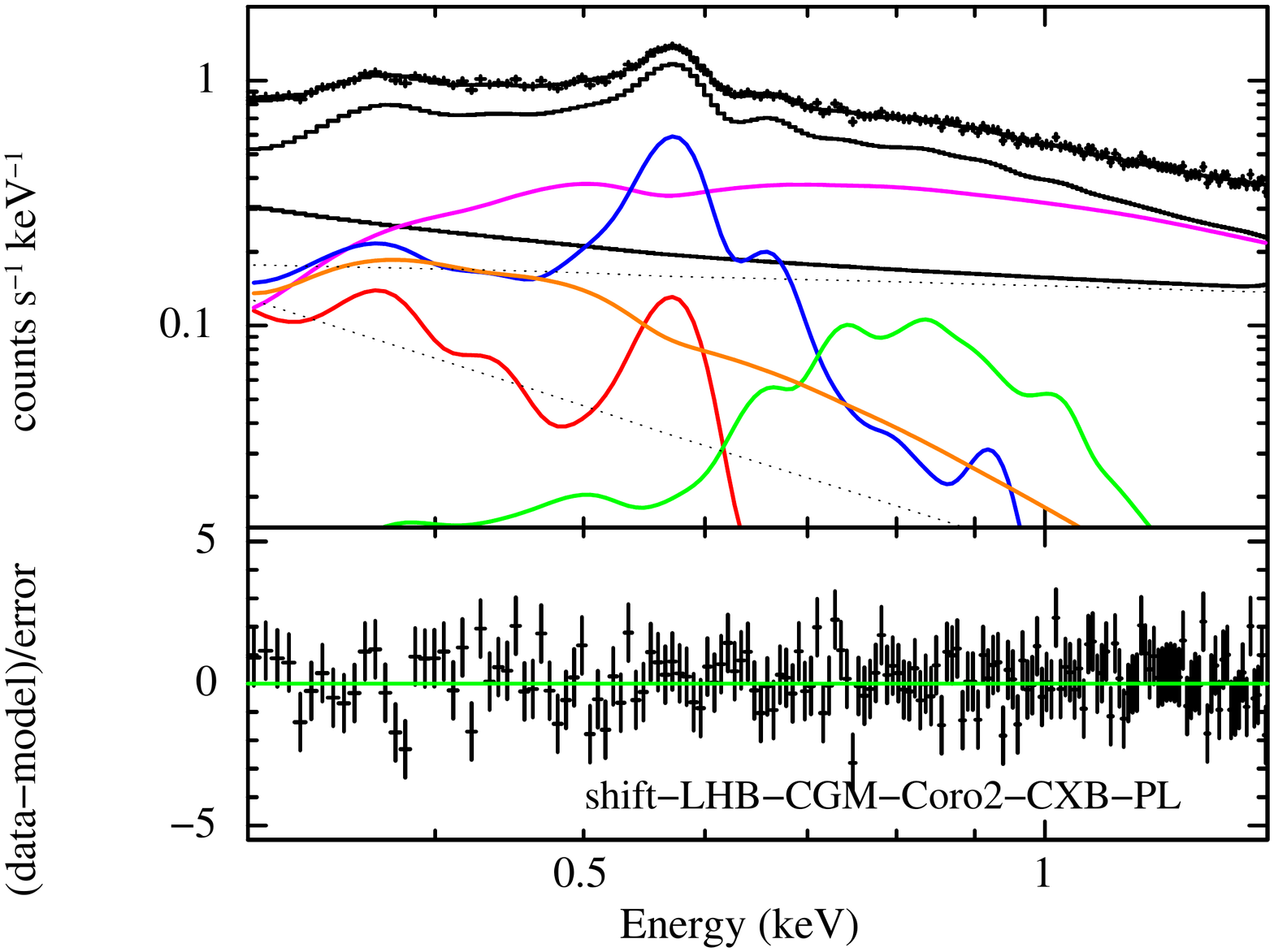}
\includegraphics[width=0.49\textwidth]{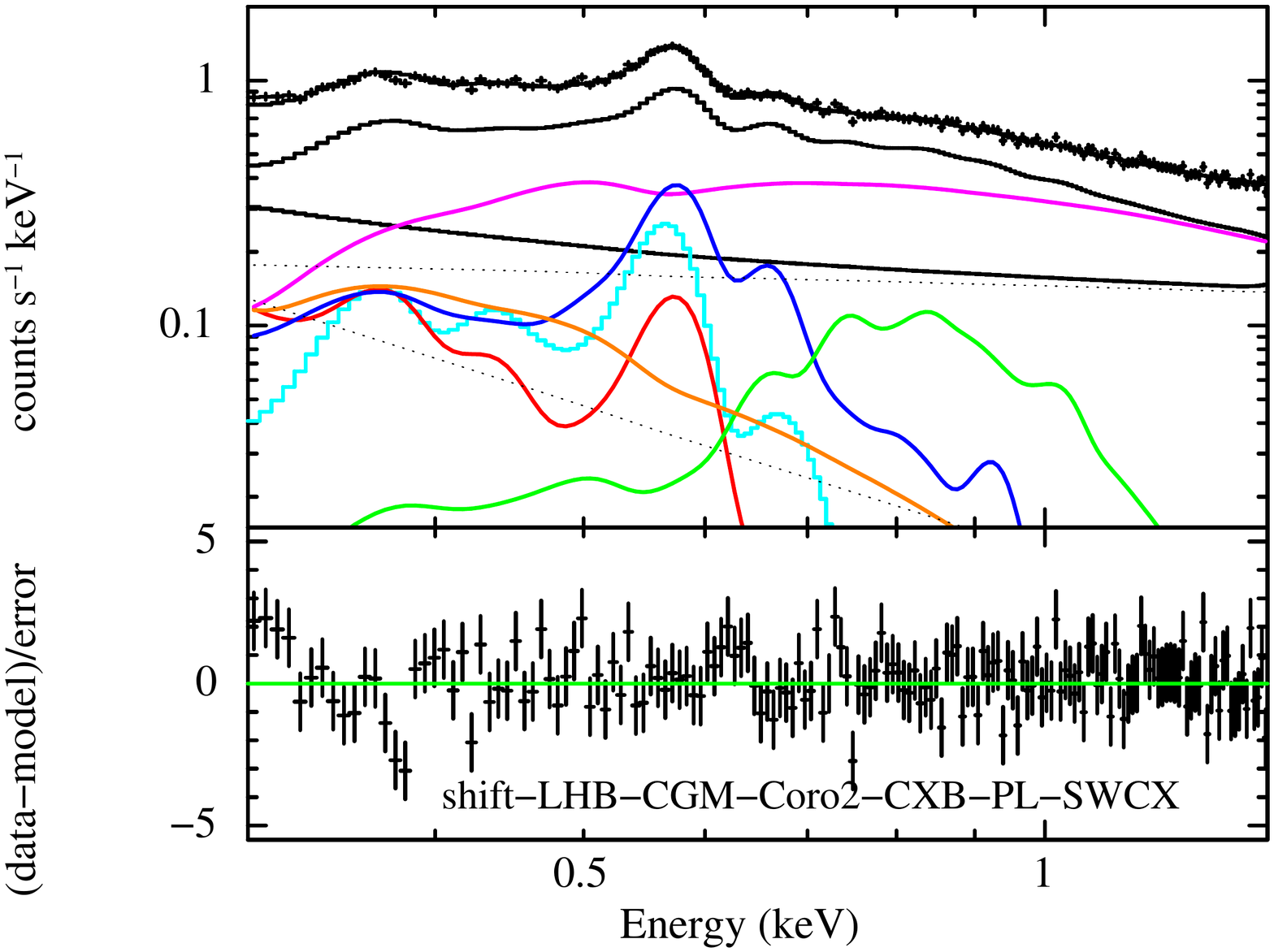}
\caption{Same spectrum and color scheme as in Fig. \ref{FCoro2}. 
{\it (Left panel)} Best fit model (shift-LHB-CGM-Coro2-CXB-PL in Tab. \ref{CGMAbun}) after the inclusion of an additional non-thermal component (power law; shown with the orange solid line) to the same spectrum and model components shown in Fig. \ref{FCoro2}.
{\it (Right panel)} Same as left panel, with the addition of the SWCX component (shift-LHB-CGM-Coro2-CXB-PL-SWCX model). }
\label{F2CXB}
\end{figure*}
Both panels of Fig. \ref{F2CXB} show that the addition of a steep power law can reproduce the bulk of the continuum emission in the soft X-ray band, therefore allowing the abundance of the CGM to be as high as $Z_{CGM}=0.3$~$Z_\odot$. 
Furthermore, both in the case of negligible and high SWCX emission, the addition of the steep power law improves the fit by $\Delta\chi^2=2.7$ and $10.8$, for the same degrees of freedom, respectively (see Tab. \ref{CGMAbun}). 
The orange lines in both panels of Fig. \ref{F2CXB} show the contribution from the hypothetical additional soft power law component. 

The best fit slope of the power law results to be extremely steep ($\Gamma=4.3\pm0.3$ and $\Gamma=4.8\pm0.4$, respectively).
Indeed, we observe that the main effect of such component is to try to mimic the emission of the continuum produced by the CGM component (Bremsstrahlung plus recombination), in order to allow higher metal abundances of the CGM (Tab. \ref{CGMAbun}). 

The slope ($\Gamma\sim4.3-4.8$) of this additional power law is too steep to be associated with a non-thermal phenomenon. 
Additionally, the lack of emission lines appears unlikely associated with a thermal phenomenon in the local Universe. 
Therefore, we conclude that either the CGM abundances are indeed as low as $Z\sim0.05-0.1$~Z$_\odot$, or such additional power law must be associated with a thermal component from the distant Universe\footnote{We also exclude that the additional power law is induced by a constant flux of soft protons interacting with the structure and detectors of \erosita.
Indeed, despite a stable-and-faint flux of soft protons might still be present also after applying the {\sc flaregti} tool, therefore potentially inducing soft X-ray emission, the same soft X-ray emission must then be observed also during the filter wheel closed observations. 
On the contrary, the additional power law component is not detected during the filter wheel closed data. }

Filaments in the Universe as well as hot baryons in the outskirts of virialised regions are expected to have temperatures lower than 1 keV and to have low abundances $Z\sim0.05-0.1$~Z$_\odot$ (Roncarelli et al. 2012; Vazza et al. 2019). 
In theory, if filaments at different redshift would contribute to the soft X-ray emission, then the emission lines associated with the thermal spectra of the filaments will appear smeared out by the redshift distribution. 
Therefore, the resulting spectrum is expected to appear as a rather steep power law with slope of $\Gamma\sim1.5$ and $\Gamma\sim3.8$ in the 0.3-0.8 and 0.8-2.0~keV, respectively (Roncarelli et al. 2012). 
Despite the slope of the additional power law appears somehow steeper than these values, we highlight the resemblance between the expected spectrum from the hot baryons in filaments at different redshifts and the power law observed here. 

Roncarelli et al. (2012), after assuming different recipes for galactic winds and black hole feedback, have estimated the surface brightness of both the whole intergalactic medium (i.e. all the gas) and of only its warm-hot component. 
They find a surface brightness for the former and the latter to be $\sim3.1-24.5\times10^{-13}$~erg cm$^{-2}$ s$^{-1}$ and $\sim0.9-3.2\times10^{-13}$~erg cm$^{-2}$ s$^{-1}$, respectively, in the 0.5-2.0~keV band. While they find, in the 0.3-0.8~keV band, a surface brightness of $\sim2.2-12.0\times10^{-13}$~erg cm$^{-2}$ s$^{-1}$ deg$^{-2}$ and $\sim1.0-3.3\times10^{-13}$~erg cm$^{-2}$ s$^{-1}$ deg$^{-2}$, respectively. 

The observed surface brightness of the best fit additional power law is $\sim1.9-3.5\times10^{-13}$~erg cm$^{-2}$ s$^{-1}$ deg$^{-2}$, in the 0.6-2.0~keV band, in the case of high and negligible SWCX contribution, respectively. 
This is within the expected range of fluxes expected from the warm-hot intergalactic medium and, as expected, it is much fainter than the total intergalactic medium emission. 
Indeed, the emission from galaxy clusters is already included in our fiducial CXB model (Gilli et al. 2007). 

On the contrary, in the 0.3-0.8 keV band, the observed surface brightness of the best fit power law emission is $\sim7.6-9.9\times10^{-13}$~erg cm$^{-2}$ s$^{-1}$ deg$^{-2}$, therefore compatible with the highest estimates for the whole intergalactic medium and a factor of $\sim3-10$ times larger than the surface brightness of only its warm-hot component. 
This indicates that either the contribution of clusters is underestimated in our fiducial CXB model or that the additional power law is physically not well justified and the CGM abundance is low. 
We also stress again that if all of the flux associated with the additional power law component is associated with the warm-hot intergalactic medium, then it would be a lucky coincidence that the power law contributes to the total spectrum in such a way to reproduce the continuum of the the CGM emission. 

To conclude, it is most likely that the warm-hot intergalactic medium contributes to the soft X-ray background, however to asses whether its contribution is strong enough to significantly affect the estimated CGM abundance, a self consistent modelling of the contributions of the different components of the extra-galactic CXB emission (i.e., AGN; galaxies; clusters; groups; warm-hot intergalactic medium; etc. ) must be performed in a self consistent way both in the data and in the simulations. 
Unfortunately, this is beyond the scopes of the current work, however this calls for a deeper understanding of the contribution of warm-hot intergalactic medium to the soft X-ray background. 

\begin{figure*}[t]
\centering
\includegraphics[width=0.49\textwidth]{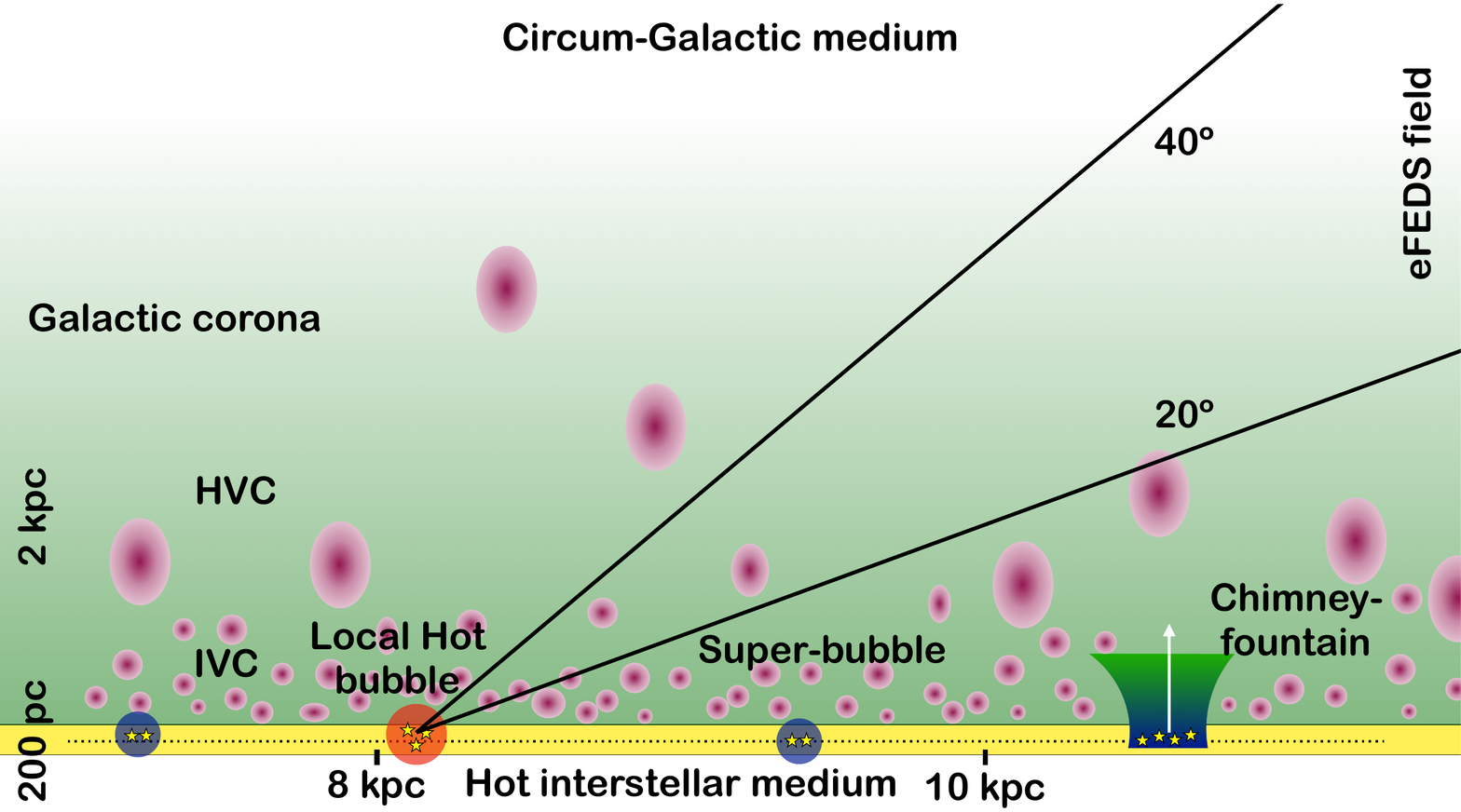}
\includegraphics[width=0.49\textwidth]{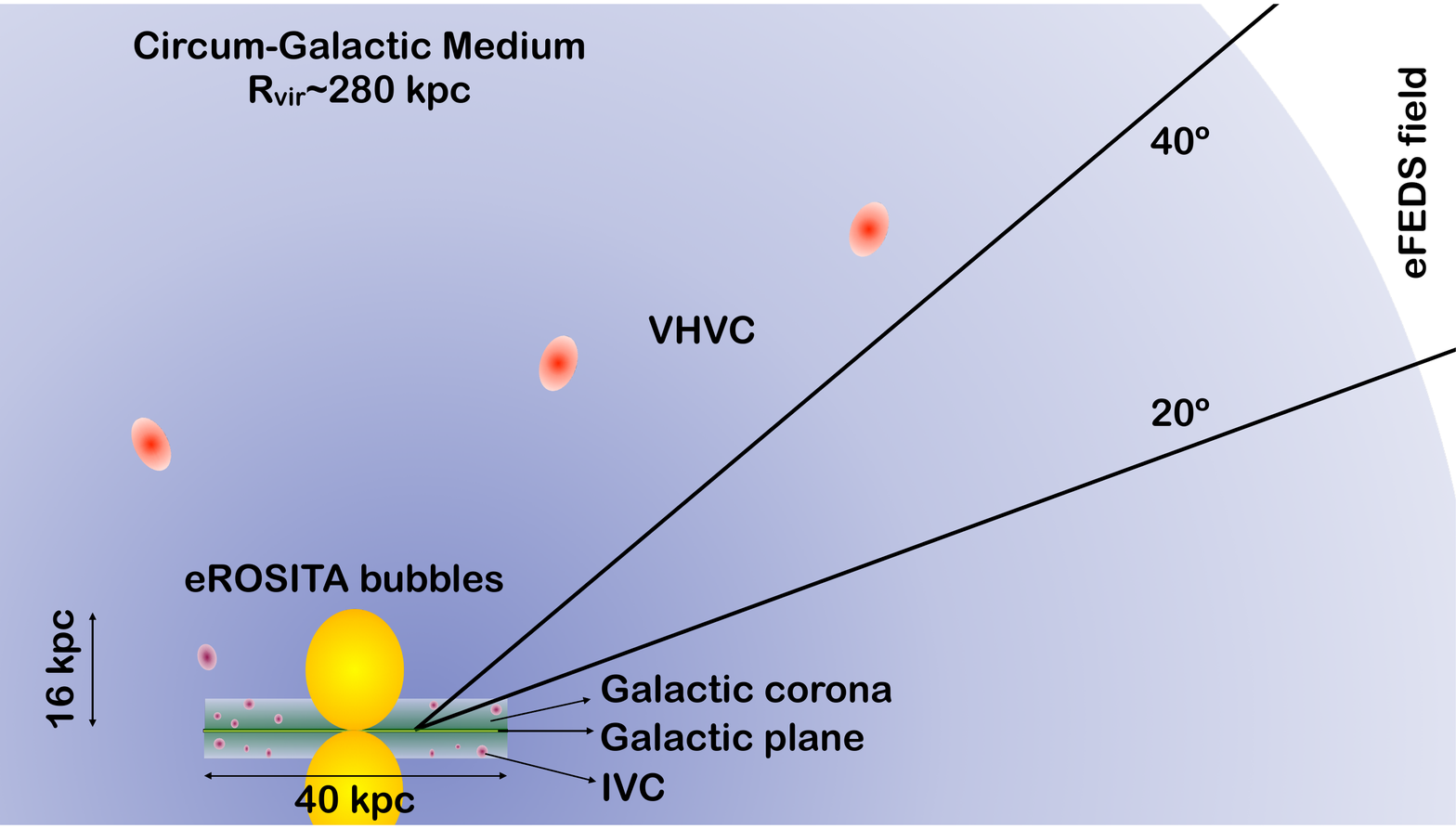}
\vspace{-0.5cm}
\caption{ Over-simplified schematic view of the different components of the diffuse emission along the line of sight towards the \efeds\ field. {\it (Left panel)} The Galactic disc is represented in yellow, with a scale-height of $\sim100$~pc. Energetic activity in the disc (yellow stars) generates copious amounts of hot plasma, which (when confined to the disc) inflates bubbles, super-bubbles (blue circles), forming features similar to the LHB (red circle). Sometimes the energetic activity has enough power to produce an outflow which breaks free into the Galactic corona, forming chimneys or fountains. Therefore, this process releases hot plasma, energy, metals and particles which energise and sustain the Galactic corona. Within the corona the intermediate and high velocity clouds (IVC and HVC, respectively) are observed, composed primarily of atomic hydrogen (red ellipses). It is likely that the intermediate velocities HI clouds represent the other phase of a cycle where hot material is expelled from the disc to then come back as cold gas. {\it (Right panel)} The blue sphere indicates the extent of the virial radius of the Milky Way (which is a proxy for the extent of the CGM) compared with the extent of the Galactic disc (assumed here to have a diameter of 40 kpc). The Galactic corona (green) is depicted above and below the Galactic disc and within it the IVC (red ellipses), while within the CGM the very high velocity clouds (VHVC) are represented (red ellipses). The yellow bipolar ellipses at the center of the disc represent the \erosita\ bubbles. }
\label{Scheme}
\end{figure*}

\subsubsection{Other biases on the observed CGM metal abundance}

Early observations of non-virialised hot plasma have found somehow lower abundances, compared with expectations, when fitted with a single temperature component (such as done here). 
Indeed, it is now recognised that plasma with a significant spread in temperature can obtain best fit values of the metal abundance which are biased towards lower values, when fitted with single temperature models. 
In theory, this might be a concern also for the CGM of the Milky Way. 
On the other hand, the indication that the CGM temperature derived from the continuum, the Oxygen lines ratio and the G-ratio of the \os\ triplet agree with each other suggests that the spread in temperature of the CGM plasma might be small enough to induce only a small bias in the best fit metal abundances measured here. However, we leave the detailed investigation of such issue for future works. 

\

We conclude that the best fit metal abundance of the CGM along the direction of the \efeds\ field is: $Z_{CGM}=0.068\pm0.004$ (statistical); $Z_{CGM}=0.052-0.072$~$Z_\odot$, once the systematic uncertainty on the contribution of the SWCX is considered and; $Z_{CGM}=0.04-0.10$~$Z_\odot$, once also the systematic uncertainties on the extrapolation of the CXB at low energy are folded in. We also note that the abundances can be as high as $Z_{CGM}=0.3$~$Z_\odot$, if the warm-hot intergalactic medium provides a contribution with a flux of $9.9\times10^{-13}$~erg~cm$^{-2}$~s$^{-1}$~deg$^{-2}$ to the soft X-ray background in the 0.3-0.6~keV band. 

\section{Discussion}

After removing the periods affected by emission from SWCX (during e0 and e3) and performing simplifying assumptions on the level of SWCX contamination during e12, we have fitted the integrated X-ray emission observed by \erosita\ in the \efeds\ field with a combination of four components: the un-absorbed emission from the LHB; the CXB; the CGM; and the Galactic corona. 
We stress that the presence of the Galactic corona, in addition to the CGM component, is indeed strictly required by the \erosita\ data. 
Additionally, we have tested the impact on our best fit results induced by a non-negligible SWCX flux during e12. 
This decomposition of the soft X-ray background is the first step towards the development of a comprehensive galaxy-CGM-corona model (Fig. \ref{Scheme}). 
In particular, a schematic picture of such interaction will be discussed in Section 9.6.

The mean surface brightness observed by \erosita\ in the \efeds\ field in the total (0.3-2~keV), the soft (0.3-0.6~keV) and the medium (0.6-2~keV) bands is: $12.6\times10^{-12}$~erg cm$^{-2}$ s$^{-1}$ deg$^{-2}$, $5.1\times10^{-12}$~erg cm$^{-2}$ s$^{-1}$ deg$^{-2}$, and $7.5\times10^{-12}$~erg cm$^{-2}$ s$^{-1}$ deg$^{-2}$, respectively (Tab. \ref{fluxes}). 

\subsection{Is a contribution due to SWCX emission during e12 required?}

When comparing the best fit models with the assumption of negligible and high SWCX fluxes, we note that the former provide significantly better fits ($\Delta\chi^2=33.6$ for the same degrees of freedom; see Tab. \ref{TCoro}). 
However, the detailed comparison of the residuals shows that the highest deviations occur in the softer band, below $\sim0.35$~keV, where a non-negligible contribution due to the electronic noise of the \erosita\ cameras is possible. 
Therefore, we do not think that this evidence can be considered as a demonstration that the SWCX emission must be negligible during e12. 
Instead, we leave such measurement for future works (which will better address the contribution of electronic noise and of the time variations of SWCX; Dennerl et al. in prep. Yeung et al. in prep.). 

\subsection{Composition of the observed background and CXB resolved fractions}
\label{Composition}

The emission in the 0.6-2~keV band is dominated by the CXB component, which alone composes more than $>83$~\% of the flux, both in the case of negligible and high SWXC contribution. 
The remaining emission is due to the Galactic corona, producing about $\sim8-9$~\% of the flux and the CGM, contributing to $\sim6-7$~\% of the total.
This is consistent with the fact that about $\sim81$~\% of the flux in the 0.5-2 keV band has been resolved into discrete sources, thanks to ultra-deep surveys with \chandra\ and \xmm\ (Luo et al. 2017). 
Additionally, this suggests that the majority of the remaining X-ray flux ($\sim15$~\%) is truly diffuse and likely due to the emission from the CGM and the Galactic corona\footnote{As it was mentioned in Sect. 7.1, a fraction of the emission that we attribute to the Galactic corona might be due to M dwarf stars. }.
However, we expect that the fractional contribution due to the Galactic emission varies greatly with Galactic latitude. 
In particular, we expect that the emission from the Galactic corona drops significantly with larger Galactic latitudes (see Locatelli et al. in prep.). 

We note that most of the ultra-deep surveys have been carried out at large Galactic latitudes ($b>50^\circ$). 
Therefore, we expect that the contribution from the Galactic corona might be smaller than the amount observed in the \efeds\ field ($\sim9$~\%) along the lines of sight investigated in such ultra-deep fields (Brandt et al. 2021).  

In the soft band (0.3-0.6~keV), one third of the flux is produced by the CXB component. 
We stress, once more, that the extrapolation of this component at energies lower than $\sim0.5-1$~keV carries significant uncertainties. 
The observed flux from the CGM, which carries most of the flux in the soft band, is highly affected by the assumed flux of the SWCX component. 
Indeed, the CGM is observed to encompass $\sim47$~\% of the flux, under the assumption of negligible SWCX, while this percentange drops to $\sim30$~\% in the case of high SWCX flux.
Considered that the SWCX emission, under the assumption of a high SWCX flux, contributes to the level of $\sim15$~\%, this is consistent with the idea that all of the flux attributed to the SWCX component is taken from the CGM one, while the flux of the LHB and of the Galactic corona appear to be nearly unaffected by the assumption on the SWCX flux. 
Indeed, the former is observed to contribute to $\sim18-19$~\% to the soft X-ray emission, while the latter has a negligible contribution of $\sim2$~\% to the soft X-ray emission. 
\begin{table}
\small
\centering
    \caption{Surface brightness of the various components of the diffuse emission. Nomenclature as in Tab. \ref{TCoro}. 
    In each couple of columns, we provide the surface brightness derived from the best fit obtained with the models shift-LHB-CGM-Coro2-CXB and shift-LHB-CGM-Coro2-CXB-SWCX, respectively (see Tab. \ref{TCoro}). 
    For the CXB-PL we report the flux of the steep power obtained with the model shift-LHB-CGM-Coro2-CXB-PL and shift-LHB-CGM-Coro2-CXB-PL, respectively (see Tab. \ref{CGMAbun}). 
    Fluxes are in units of $10^{-13}$ erg cm$^{-2}$ s$^{-1}$ deg$^{-2}$. }
    \label{fluxes}
    \begin{tabular}{l | r r | r r | r r}
    \hline \hline
    \multicolumn{7}{c}{\bf Surface brightness} \\
    \hline \hline
Energy & \multicolumn{2}{c}{0.3-2.0} & \multicolumn{2}{c}{0.3-0.6} & \multicolumn{2}{c}{0.6-2.0} \\
       & (keV)      & (keV)          & (keV)     & (keV)           & (keV)  & (keV)  \\
       &            & SWCX           &           & SWCX            &        & SWCX   \\
LHB    & $9.5$      & $10.1$         & $9.2$     & $9.8$           & $0.3$  & $0.3$  \\
CGM    & $29.7$     & $20.5$         & $24.1$    & $15.6$          & $5.6$  & $4.9$  \\
Coro2  & $7.4$      & $7.7$          & $1.0$     & $1.1$           & $6.4$  & $6.6$  \\
CXB    & $79.8$     & $79.5$         & $17.1$    & $17.0$          & $62.8$ & $62.6$ \\ 
SWCX   &            & $8.2$          &           & $7.6$           &        & $0.7$  \\
\hline
Total  & \multicolumn{2}{c}{$126.4$} & \multicolumn{2}{c}{$51.4$}  & \multicolumn{2}{c}{$75.1$} \\
\hline
CXB-PL & $13.4$     & $9.5$          & $9.9$     & $7.6$           & $3.5$ & $1.9$ \\ 
\hline
\end{tabular}
\end{table}

We have shown that, if the CGM abundances is $Z_{CGM}=0.3$~$Z_\odot$, then a steep ($\Gamma=4.3-4.8$) power law component (likely associated with the warm-hot intergalactic medium) is required to fit the diffuse emission from the \efeds\ region. Such component must have a flux of $F_{PL}\sim(7-10)\times10^{-13}$~erg cm$^{-2}$ s$^{-1}$ deg$^{-2}$ in the 0.3-0.6~keV band, therefore contributing to $\sim15-19$~\% to the soft X-ray background. 

\subsection{"A posteriori" validation of the best fit}

We observe that the parameters of our best fit models\footnote{We consider here: shift-LHB-CGM-Coro2-CXB; shift-LHB-CGM-Coro2-CXBs; shift-LHB-CGM-Coro2-CXB-PL; as well as their versions considering the SWCX contribution shift-LHB-CGM-Coro2-CXB-SWCX; shift-LHB-CGM-Coro2-CXBs-SWCX and; shift-LHB-CGM-Coro2-CXB-PL-SWCX.} show a small dependence on the extrapolation of the CXB component below $\sim1$~keV. 
Indeed, when comparing the final best fit models, we observe that all the best fit parameters are consistent within each other, apart from the temperature, normalisation and metal abundance of the CGM\footnote{As a consequence of the assumption on the CXB shape, then the normalisation of the CXB emission are different in the two models, however they are consistent with the expected values of 0.269 photons s$^{-1}$ cm$^{-2}$ at 1 keV and 0.34 photons s$^{-1}$ cm$^{-2}$ at 1 keV for the CXB and CXBh models, respectively. This is not surprising. Indeed, the CXB component is anchored by the data at high energy, therefore a different assumption on its extrapolation in the soft X-ray band, then induces a different normalization of the CXB component at 1 keV. } (see Tab. \ref{TCoro} and \ref{CGMAbun}). 

When comparing in detail each single best fit model with its version including SWCX, we observe that the addition of the SWCX component induces a significantly hotter CGM and a correspondingly lower CGM normalisation. 
In particular, consistent temperatures are measured for models considering either negligible or high SWCX contribution. 
Additionally, as detailed in Sect. \ref{Composition}, for each model the addition of the SWCX component decreases the flux attributed to the CGM component, making its normalisation to drop (see Tab. \ref{TCoro}, \ref{CGMAbun} and \ref{fluxes}). 

We observe that, despite being a free parameter of the model, the normalisation of the CXB and CXBs components are consistent with the expectations. 
Also the normalisation of the LHB component is consistent, within 1.5 sigmas ($N_{LHB}=0.0032\pm0.0005$ and $0.0034\pm0.0006$~pc~cm$^{-6}$ for the shift-LHB-CGM-Coro2-CXB and shift-LHB-CGM-Coro2-CXB-SWCX models, respectively) with its expected value ($N_{LHB} = 2.7\times10^{-3}$~pc~cm$^{-6}$). 
This, therefore, verifies "a posteriori" that the soft X-ray flux attributed to the LHB component is consistent with the flux observed by \rosat\ along the same line of sight (Liu et al. 2017). 
As specified in Sect. \ref{CXBs} and \ref{AddPL}, for the shift-LHB-CGM-Coro2-CXBs and shift-LHB-CGM-Coro2-CXB-PL models (and their versions including the SWCX component), we observed that the initial best fit was finding a normalisation of the LHB component which would produce a flux in the R1 and R2 bands significantly larger than the one observed by \rosat. 
For these reasons, we fixed the LHB normalisation in those models. 

\subsection{Temperature of the CGM}

By fitting the spectrum of the \efeds\ field, we constrained the temperature of the CGM to a very small statistical uncertainty of the order of $\sim3$~\% (Tab. \ref{TCoro}). 
We measured that the poorly constrained SWCX contribution during e12 would induce a variation in temperature from $kT_{CGM}=0.157\pm0.004$~keV to $kT_{CGM}=0.173\pm0.005$~keV for negligible and high flux, respectively, therefore a systematic uncertainty as large as $\sim10$~\%. 

Assuming the validity of the virial theorem and that the gravitational potential is dominated by a Navarro, Frenk \& White (1997) profile of the dark matter halo, then theoretical arguments suggest that the CGM should be iso-thermal (although we know that this is not true for galaxy clusters). 
In particular, assuming the validity of the equation: $kT_{vir} = \frac{G M_{vir} \mu m_p}{2 R_{vir} }$ (where $G$ is the gravitational constant, $\mu$ is the mean atomic number per atom, $m_p$ is the proton mass), we can estimate an order of magnitude for the virial temperature\footnote{Different assumptions (e.g., regarding the presence of pressure at the virial radius, etc.) can lead to expected virial temperatures which differ by $\sim30$~\% or more. }. 

We compute the mean atomic number per atom which corresponds to the solar abundances assumed here (Lodders et al. 2003). Indeed, for all the elements X heavier than He, the solar abundance $z(X)$ value is rescaled for a constant factor $\zeta$ to fit the data. The mean atomic number per atom $\mu$ can then be derived as $\mu = 1 + 2 * z(He) + \sum_X 2 * z(X) * \zeta * \xi(X)$ where $z(X)$ is the atomic number of the element $X$. For $\zeta \sim 0.1$ and the assumed solar abundance values, we obtain $\mu = 1.32$.
Then, by assuming a virial mass and radius of $M_{vir}=1.3\pm0.3\times10^{12}$~M$_\odot$ and $R_{vir}=282\pm30$~kpc, respectively (Bland-Hawthorn \& Gerhard 2016), we obtain $kT_{vir}\sim0.14$, which considering the uncertainty on $M_{vir}$ results in $kT_{vir}=0.14\pm0.04$~keV, which is consistent with the observed value. 

It is reassuring that looking outwards, towards directions where the influence of the dark matter is expected to be larger, the CGM temperature corresponds to the virial prediction. However, this assumption breaks down close to the Galactic disc, where the CGM plasma is heated to form the Galactic corona (as this work demonstrates) as well as close to the Galactic center, where the \erosita\ bubbles and Galactic center chimneys testify the presence of significantly hotter plasma (Ponti et al. 2019; 2021; Predehl et al. 2020). 

We also note that other studies, looking towards other lines of sight, measured different temperatures of the CGM (Henley et al. 2010; 2015; Miller \& Bregman 2013; 2015; Yoshino et al. 2009; Nakashima et al. 2018; Kataoka et al. 2018). Therefore, it is most likely that the real scatter on the CGM temperature is significantly larger than our statistical uncertainty. 

\subsection{CGM abundances}

We measured the abundance of the hot CGM with a statistical uncertainty of $\sim5$~\%. 
We note that the systematic uncertainty on the abundances, induced by the SWCX contribution during e12, is of the order of $\sim15$~\%, therefore the abundance is estimated to be within the range from $Z_{CGM}=0.058\pm0.006$ $Z_\odot$ to $Z_{CGM}=0.068\pm0.004$ $Z_\odot$ (Tab. \ref{TCoro}).
Additionally, we showed that the uncertainty on the proper extrapolation below $1.4$~keV of the CXB component results in a larger systematic uncertainty of the order of $\sim70$~\%, with the abundance within the range $Z_{CGM}=0.04-0.10$ $Z_\odot$.
Such values are several times lower than the value ($Z_{CGM}=0.3$ $Z_{\odot}$) typically assumed to model the observed emission and absorption lines induced by the CGM (see Bregman et al. 2018 and references therein). 
Therefore, we expect that such difference will have a significant impact on some of the derived parameters of the hot CGM, such as its density and total mass. 

The observed abundances, despite being lower than typically assumed, are in good agreement with predictions from cosmological simulations, which forecast an abundance of $\sim0.1$~solar or lower for the outer hot CGM in Milky Way galaxies in the present day Universe (Crain et al. 2010). 
Indeed, cosmological simulations suggest that the CGM gets enriched with metals within a central funnel, where outflows can produce features such as the \erosita\ bubbles (Predehl et al. 2020; Pillepich et al. 2021). 
Additionally, the energetic activity within the Galactic disc is expected to drive super-bubbles, fountains and chimneys which collectively are sustaining a metal enriched corona present just above the Galactic disc, such as the one observed in this work. 
On the other hand, the outer CGM, away from the central funnel, appears to be less affected by outflows and feedback, therefore its metal abundance is closer to the one in the filaments of the large scale structure and the pristine composition. 

Is the metal abundance measured here in agreement with independent constraints? 
The line of sight towards the Large Magellanic Cloud allows us to estimate the metal abundance towards that direction by combining the electron column density derived from pulsar dispersion measure and the equivalent width of the \os\ line (Wang et al. 2005; Yao et al. 2009; Fang et al. 2013; Miller \& Bregman 2013; 2015; Miller 2016). 
Assuming the current best fit model for the density distribution within the CGM, several authors have estimated the CGM abundances being larger than 0.3~solar (see Bregman et al. 2018 and reference therein). 
We point out that the current "state-of-the-art" models of the density of the CGM might be affected by the hot plasma associated with the \erosita\ bubbles (Locatelli et al. in prep.). 
Indeed, those models have been developed before the discovery of the \erosita\ bubbles and are biased. 
Indeed, despite those works avoided line of sights passing through the \fermi\ bubbles, they contain lines of sight through the \erosita\ bubbles, which might bias the density distribution of the CGM in the center. 
Additionally, a fraction of the electrons contributing to the dispersion measure might be associated with the Galactic corona, which still misses a solid model for its density distribution (but see Kaaret et al. 2020). 
Therefore, until a detailed understanding of such effects is properly taken into account, we can not conclude that the CGM abundances must be higher than 0.3 solar between us and the LMC or whether it can be consistent with the value observed in this work of $Z_{CGM}\sim0.04-0.1$~$Z_\odot$, such as observed towards the \efeds\ field. 

\subsubsection{A metal rich CGM implies the detection of the warm hot intergalactic medium}

We have shown that, by postulating the presence of an additional non-thermal component (a power law), it is possible to recover CGM abundances as high as $Z_{CGM}\sim0.3$~$Z_\odot$ (Sect. \ref{AddPL}). 
Additionally, the properties of such component are somehow reminiscent of the expected emission from the warm-hot intergalactic medium. 
We conclude that, if it will be demonstrated that the CGM has to be metal rich ($Z_{CGM}\gg0.3$~$Z_\odot$), then a significant fraction ($\sim15-19$~\%) of the soft X-ray background is required to be produced by the warm-hot intergalactic medium. 

\subsection{A schematic view of the hot phase of the Milky Way}

Figure \ref{Scheme} shows a schematic view of the different components to the diffuse soft X-ray emission observed in the \efeds\ field. 
The yellow stripe along the Galactic plane aims to represent the Galactic disc, where the multi-phase ISM is laying.  
Therefore, we represent such component with a thick stripe along the Galactic disc with a scale-height of $h=\pm100$~pc, which corresponds to the scale-height of the cold and molecular phase (Ferriere 2001). 

The Sun is located within such stripe about $\sim24$~pc above the mid plane of the Milky Way (Bland-Hawthorn \& Gerhard 2016). 
The emission associated with the SWCX is expected to be produced within several Astronomical Units, therefore it appears too small to be drawn in Fig. \ref{Scheme}. 
For this reason, we omit it. 

The red circle in Fig. \ref{Scheme} represents the emission from the local hot bubble, which is characterised by a radius of $\sim200$~pc and it contains the Sun close to its center (Liu et al. 2017).  
It is expected (and observed) that the shape of the local hot bubble deviates significantly from the perfect sphere displayed in Fig. \ref{Scheme} as a consequence, for example, of larger resistance to its expansion encountered on the Galactic plane (Liu et al. 2017; Zucker et al. 2022).  

We note that, for lines of sight spanning Galactic latitudes within $20<b<40^\circ$ (such as for the \efeds\ field), we do not expect to observe a large contribution from the hot phase of the ISM, in addition to the emission from the LHB (for an ISM scale-height of $h=\pm100$~pc; see Fig. \ref{Scheme}). 
On the other hand, the geometry represented in Fig. \ref{Scheme} indicates that the emission observed in the \efeds\ field should be rather sensitive to any extended hot atmosphere of the Milky Way, with a scale-height of few kpc, such as a Galactic corona, as shown by simulations of hot interstellar medium (Kim \& Ostriker 2017). 
Therefore, it is not surprising that, thanks to this study, we have revealed and singled out both the Galactic corona and the Galactic hot halo (CGM). 
The right panel of Fig. \ref{Scheme} sketches both the CGM (in blue), assumed to extend to the virial radius, and the Galactic corona (in green, sandwiching the Galactic disc, assumed to have a 40 kpc diameter). 

Given the low temperature of the CGM of the Milky Way, it is expected that rapid cooling will characterise the coolest regions, if temperature fluctuations are present. 
The CGM plasma within these cool pockets will then rapidly radiate their energy and condense to form low temperature and high density gas, which might appear as UV-optical line emitter-absorbers or as HI clouds, at several tens of kiloparsecs from the Milky Way. 
If so, then a tight correlation between the physical properties of the hot Galactic phase (i.e., CGM and Galactic corona) and the colder ones must be present. 
The Milky Way contains a significant amount of extra-planar cold material moving at high speed through the CGM. 
Such cold (HI) clouds have been arbitrarily divided into very high, high and intermediate velocity clouds (VHVC, HVC and IVC, respectively)\footnote{ We define as IVC HI clouds with line of sight velocities within $40\leq |v_{LSR}| \leq90$ km s$^{-1}$, then HVC HI clouds with $90\leq |v_{LSR}| \leq170$ km s$^{-1}$ and very high velocity clouds HI clouds with  $ |v_{LSR}| > 170$ km s$^{-1}$.} on the base of their observed speed. 
The physical origin of such clouds is still debated and some argue that all clouds share the same origin with the VHVC and HVC being more extreme examples of the IVC. 
Others point to a possible different origin with the VHVC and HVC being associated with stripping and-or accretion of satellite galaxies, condensation from the hot CGM or accretion from the intergalactic medium, while the IVC being associated with gas launched from the activity in the disc which then recondenses, therefore forming a Galactic fountain (Fraternali et al. 2015; Marasco \& Fraternali 2017; Putman et al. 2012; Gronke \& Oh 2020).

The scenario depicted in Fig. \ref{Scheme} predicts that the colder phases are affected by the hot-volume-filling plasma which composes the hot CGM and corona. 
If so, then we would expect the VHVC and HVC to be distributed on a significantly larger scale and to have smaller metal abundances, while the IVC to be closer to the disc and have nearly solar abundances. 
The current best estimates of the intermediate velocity clouds suggest that they possess abundances close to solar (Wakker et al. 2001, 2008) and are distributed $<1.5$ kpc above the plane of the Milky Way and have high covering factors $f_c\sim0.9$ (Putman et al. 2012; Lehner et al. 2022; Marasco et al. 2022). 
This evidence is indeed in line with the idea that the hot Galactic corona is the volume-filling plasma in which such clouds are immersed (left panel of Fig. \ref{Scheme}). 
On the other hand, the abundances of the VHVC and HVC are within the range of $\sim0.1-0.3$~solar (Wakker et al. 1999b; Fox et al. 2010; Shull et al. 2011; Richter et al. 2013; Collins et al. 2007; Richter 2017), therefore a large fraction of such clouds have significantly higher metallicities than the ones of the hot CGM ($Z_{CGM}\sim0.06$~$Z_\odot$). 
This difference might appear rather surprising, if it is assumed that all high velocity clouds are the product of re-condensation from the hot CGM. 
On the other hand, such difference might be reconciled with the picture proposed in Fig. \ref{Scheme}, if we consider that part of such clouds might originate from plasma having even higher metallicities, such as the IVC, which then is significant mixing with the metal poor hot CGM phase observed in X-rays (Heitsch et al. 2022; Marasco et al. 2022). 
Indeed, the gas stripped from satellites, large scale Galactic outflows, among other processes, might be enriched in metals, therefore have initial metallicities larger than the hot CGM phase. 

The interplay between the hot (CGM and corona) and cold (VHVC, HVC and IVC) phases of the Galaxy are important to understand galaxy evolution.  
Indeed, the dense and cold HI plasma is observed to be primarily accreted onto the Galactic disc and to be used as material to form stars and to grow the galaxies. 
We note that, as a consequence of the higher temperature of the plasma within the Galactic corona ($kT\sim0.4-0.7$~keV), the process of rapid cooling is less likely to spontaneously occur directly from the hot phase. 
However, the plasma in the Galactic corona is likely to be multi-phase and the interfaces between the colder and the hot phases might work as the seeds where rapid cooling occurs. 
Additionally, fountains models posit that adiabatic expansion might rapidly cool a good fraction of the hot outflow, which, whenever it can not escape the system, then goes back to the disc in the form of, for example, IVC. 

All these models assume that the hot and colder phases are in nearly pressure equilibrium. 
Unfortunately, we can not test this essential point in this work. 
In future works, by developing a model of the density distribution within the CGM and corona, we expect to be able to verify whether this is in agreement with the full extent of the \erosita\ all-sky survey data. 

We have achieved this thanks to the good energy resolution, the stable and low level of instrumental background in the soft ($\sim0.3-1.4$~keV) X-ray band, as well as the outstanding statistics provided by the \erosita\ spectrum of the \efeds\ field. The latter point is the result of the combination of the unprecedented grasp of the \erosita\ telescope, as well as the large sky area, the relatively deep exposure and the intermediate Galactic latitudes of the \efeds\ field.
The good energy resolution of \erosita\ have been essential to spectrally discriminate the emission from the Galactic corona from the other components. 
Indeed, the coronal plasma possesses a temperature ($kT\sim0.4-0.7$) which is incompatible with the ones associated with either the local hot bubble ($kT\sim0.1$~keV) or the CGM ($kT\sim0.15-0.17$~keV).

\section{Acknowledgments}

We thank Mattia Sormani for helpful discussion and the referee for the comments, which significantly improved the paper. 
This work is based on data from \erosita, the soft X-ray instrument aboard SRG, a joint Russian-German science mission supported by the Russian Space Agency (Roskosmos), in the interests of the Russian Academy of Sciences represented by its Space Research Institute (IKI), and the Deutsches Zentrum f\"ur Luft- und Raumfahrt (DLR). The SRG spacecraft was built by Lavochkin Association (NPOL) and its subcontractors, and is operated by NPOL with support from the Max Planck Institute for Extraterrestrial Physics (MPE).
The development and construction of the \erosita\ X-ray instrument was led by MPE, with contributions from the Dr. Karl Remeis Observatory Bamberg \& ECAP (FAU Erlangen-Nuernberg), the University of Hamburg Observatory, the Leibniz Institute for Astrophysics Potsdam (AIP), and the Institute for Astronomy and Astrophysics of the University of T\"ubingen, with the support of DLR and the Max Planck Society. The Argelander Institute for Astronomy of the University of Bonn and the Ludwig Maximilians Universit\"at Munich also participated in the science preparation for \erosita.
The \erosita\ data shown here were processed using the eSASS software system developed by the German \erosita\ consortium.

This project acknowledges funding from the European Research Council (ERC) under the European Union’s Horizon 2020 research and innovation programme (grant agreement No 865637). SB acknowledges financial support from the Italian Space Agency under grant ASI-INAF I/037/12/0 and from the PRIN MIUR project: ‘Black Hole winds and the Baryon Life Cycle of Galaxies: the stone-guest at the galaxy evolution supper’, contract number 2017PH3WAT.

\appendix

\section{The effects of assumed abundances} 
\label{SecAbu}

\begin{figure}[th]
\centering
\vspace{-1.0 cm}
\includegraphics[width=0.5\textwidth]{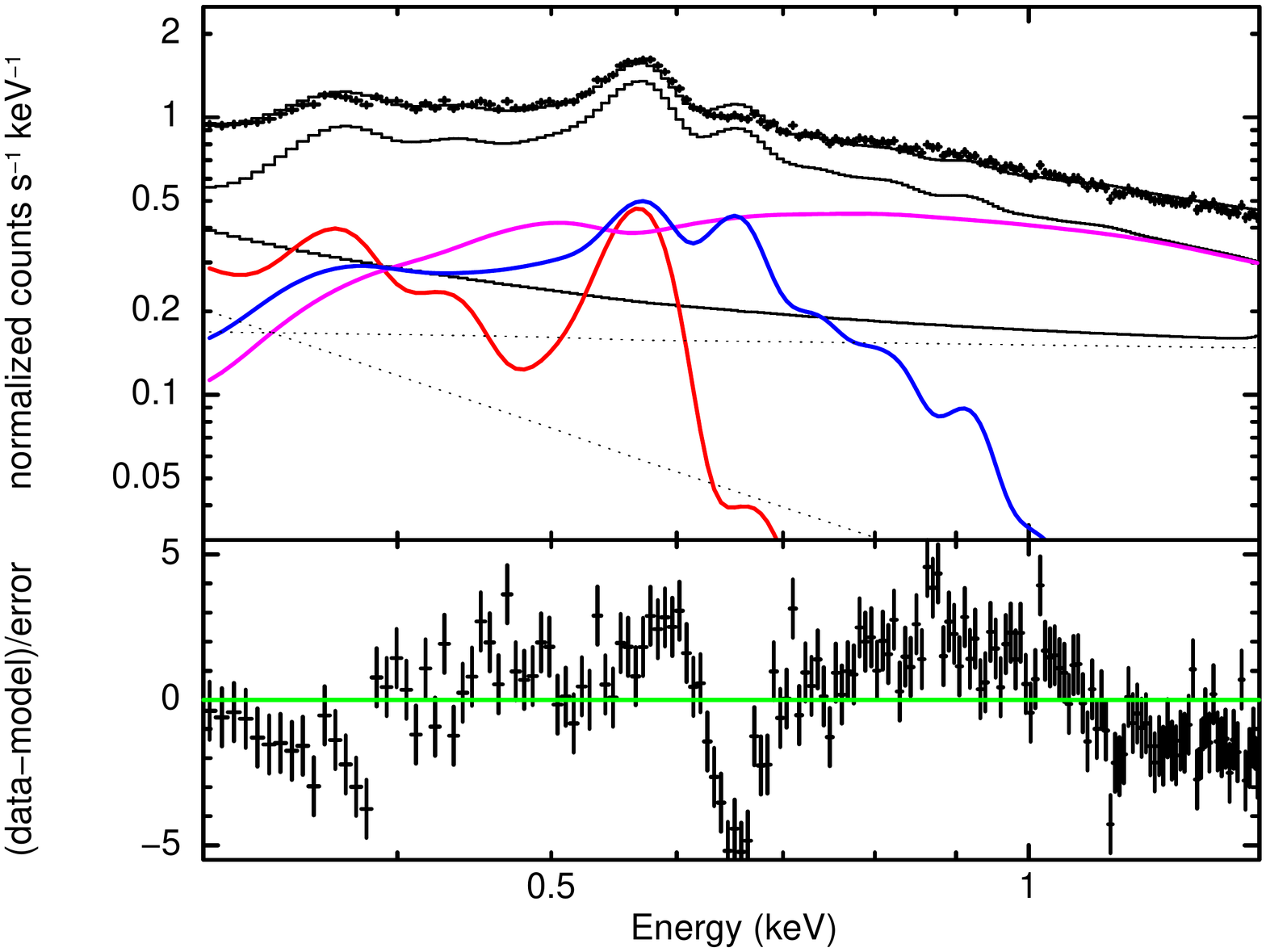}

\vspace{-1.0 cm}
\includegraphics[width=0.5\textwidth]{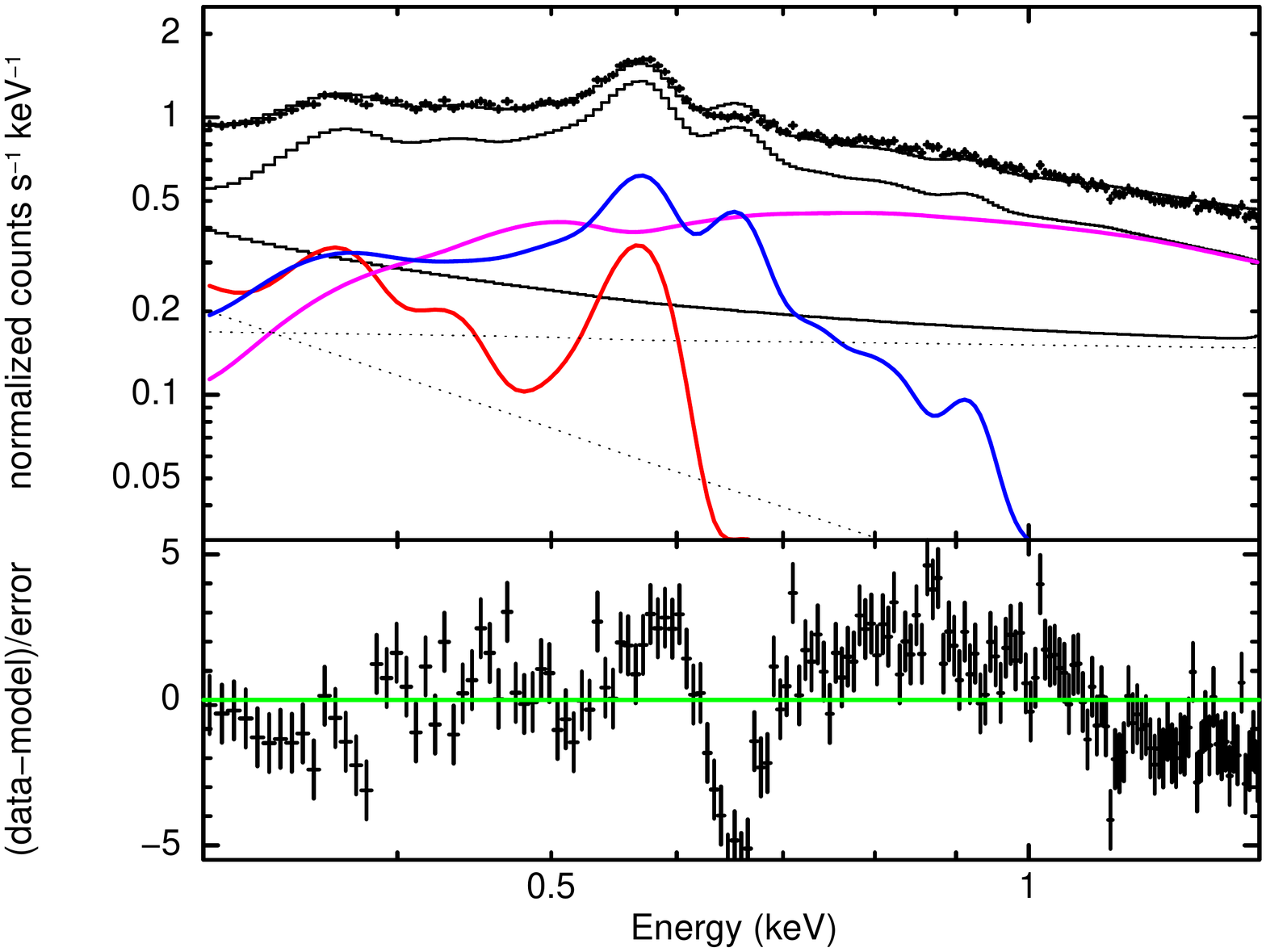}

\vspace{-1.0 cm}
\includegraphics[width=0.5\textwidth]{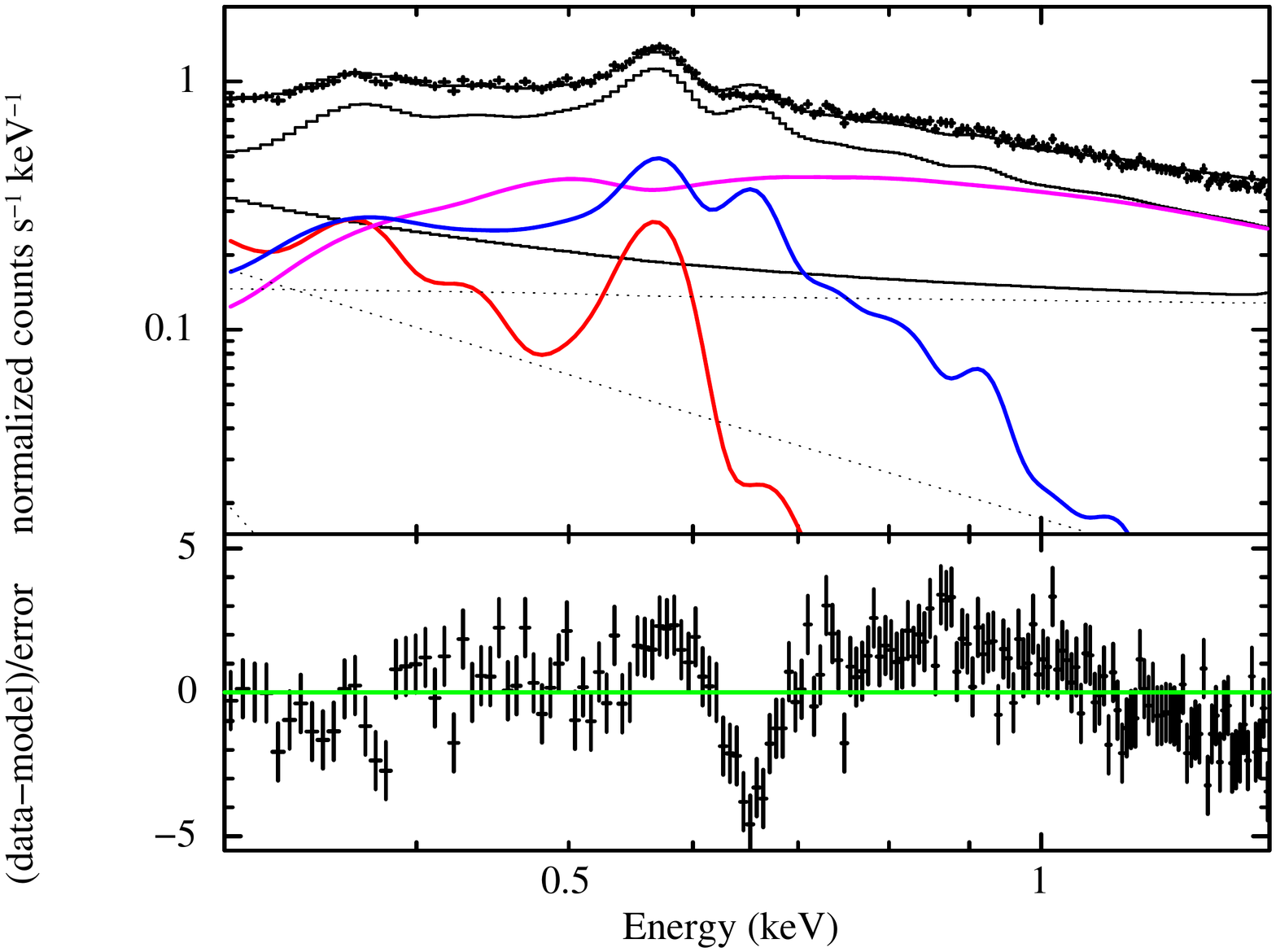}

\vspace{-0.3 cm}
\caption{Spectrum fitted with a 3 components models (LHB+CGM+CXB) assuming different metal abundances. 
The red, blue, magenta and black solid lines show the contribution from the local hot bubble, the circum-Galactic medium, the cosmic X-ray background and instrumental background, respectively. The dotted lines show the various contributions to the instrumental background. 
From top to bottom: Anders \& Grevesse (1989); Wilms et al. (2000) and Lodders (2003) abundances are assumed. 
Assuming the Anders \& Grevesse (1989) abundances, about half of the \os\ line flux is due to the local hot bubble, while for later abundances the contribution to the \os\ line drops significantly. Large residuals are present at the energy of the \oe\ line and between $\sim0.7$ and $1$ keV. }
\label{FAbu}
\end{figure}

The different panels of Fig. \ref{FAbu} illustrate the effects of changing the assumed metal abundances. 
Going from top to bottom the Anders \& Grevesse (1989), Wilms et al. (2000) and Lodders (2003) metal abundances are assumed, respectively. 
We observe that about half of the \os\ line flux is due to the local hot bubble, once the Anders \& Grevesse (1989) abundances are assumed. 
On the other hand, a lower fraction of the \os\ line emission is due to the local hot bubble, once more recently measured solar abundances are assumed. 

We note that the Oxygen, the Nitrogen and the Neon abundances drop by a factor of $\sim1.7$ from Anders \& Grevesse (1989) to both the Wilms et al. (2000) and Lodders (2003) ones. 
On the other hand, the drop of the Carbon abundance from the Anders \& Grevesse (1989) to the more recently measured ones is only of a factor of $\sim1.48$, therefore significantly smaller than the ones of the other metal lines dominating the emission in the soft X-ray band (e.g., O, N, Ne, etc.). 
In the current fits, the normalisation of the LHB component adjusts itself in order to reproduce the \cs\ emission line. 
Therefore, as a consequence of the lower metals to Carbon fraction observed in Lodder et al. (2003), the contribution of the LHB component to the \os\ line emission drops accordingly, with a larger fraction of flux instead being produced by the CGM component (where the \cs\ emission line is suppressed by the Galactic absorption), compared with the Anders \& Grevesse (1989) ones (see Fig. \ref{FAbu}). 

Another consequence of the drop in the solar metal abundances is behind the drastic change in the best fit abundances of the CGM component, which changes from $\sim0.04$ to $\sim0.07-0.08$ solar (see Tab. \ref{TAbu}). 
\begin{table}
\tiny
\centering
    \caption{Best fit parameters obtained by fitting the e12 spectrum with different models.
    Same nomenclature as for Tab. \ref{TCoro}. 
    The first, second and third columns show the best fit results, once the metal abundances from Anders \& Grevesse (1989), Lodders (2003) and Wilms et al. (2000) are assumed, respectively. }
    \label{TAbu}
    \begin{tabular}{c c c c c c c c c c c c }
    \hline \hline
    \multicolumn{4}{c}{\bf SPECTRUM e12} \\
    \hline \hline
             & AnGr             & Lod3            & Wilm             \\
             & LHB-CGM-CXB      & LHB-CGM-CXB     & LHB-CGM-CXB      \\
$N_{LHB}$    & $5.2\pm0.2$      & $4.4\pm0.2$     & $6.6\pm0.2$    \\
$N_{CXB}$    & $0.404\pm0.002$  & $0.412\pm0.002$ & $0.41\pm0.02$    \\
$kT_{CGM}$   & $0.210\pm0.003$  & $0.184\pm0.002$ & $0.196\pm0.002$  \\
$Z_{CGM}$    & $0.044\pm0.003$  & $0.072\pm0.004$ & $0.078\pm0.005$  \\
$N_{CGM}$    & $29.2\pm1.7$     & $46\pm2$        & $36\pm2$         \\
$\chi^2$     & 1414.9           & 1519.3          & 1434.1           \\
$dof$        & 815              & 815             & 815              \\
\hline
\end{tabular}
\end{table}

\section{Background flares}
\label{flareGTI}

\begin{figure}[t]
\centering
\includegraphics[width=0.49\textwidth]{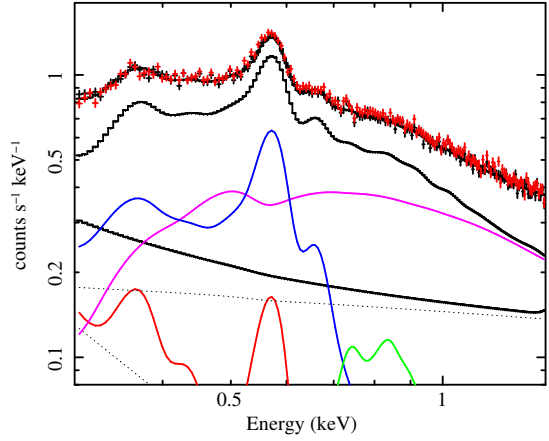}
\vspace{-0.5cm}
\caption{The black data show the e12 spectrum fitted with the best fit model shift-LHB-CGM-Coro2-CXB (same spectrum and color scheme as in Fig. \ref{FCoro}). 
The red data show the e12 spectrum obtained after filtering the events with the {\sc flareGTI} tool. The two spectra are consistent with each other. Indeed, notable background flares are not detected during the observation of the \efeds\ region during e1 and e2. }
\label{FlareGTI}
\end{figure}

As discussed in Sect. \ref{data}, we investigated the temporal evolution of the particle background during the \erosita\ observations of the \efeds\ field. 
No notable flare was detected during e1 and e2. 
The black spectrum in Fig. \ref{FlareGTI} shows the e12 spectrum investigated in this work and the best fit shift-LHB-CGM-Coro2-CXB model. 
The red spectrum shows the same spectrum, after filtering the events for background flares with the {\sc flareGTI} tool. 
The consistency between these spectra corroborate the fact that important background flares do not affect the e12 spectrum, therefore they do not have an effect on the results obtained here. 

\end{document}